\documentclass[aps,prd,onecolumn,notitlepage,nofootinbib,showpacs,preprintnumbers]{revtex4-1}
\usepackage[T1]{fontenc}
\usepackage[utf8]{inputenc}
\usepackage{bm}
\usepackage{amsmath}
\usepackage{amssymb}
\usepackage{esint}
\usepackage{color}
\usepackage{graphicx}% Include figure files
\PassOptionsToPackage{normalem}{ulem}
\usepackage{ulem}
\usepackage{array}
\usepackage{multirow}
\usepackage{hyperref}

\begin{document}
	
	\title{Propagation of the gravitational waves in a cosmological background}
	
	\author{Xian Gao}%
	\email[Email: ]{gaoxian@mail.sysu.edu.cn}
	
	\author{Xun-Yang Hong}%
	\email[Email: ]{hongxy5@mail2.sysu.edu.cn}
	
	\affiliation{School of Physics and Astronomy, Sun Yat-sen University, Guangzhou 510275, China}
	
	\date{June 17, 2019}
		
	\begin{abstract}
		We investigate the propagation of the gravitational waves in a cosmological background. Based on the  framework of spatially covariant gravity, we derive the general quadratic action for the gravitational waves. The spatial derivatives of the extrinsic curvature and the parity-violating terms are systematically introduced. Special attention is paid to the propagation speed of the gravitational waves. We find that it is possible to make the two polarization modes propagate in the same speed, which may differ from that of the light, in the presence of parity-violating terms in the action. In particular, we identify a large class of spatially covariant gravity theories with parity violation, in which both the polarization modes propagate in the speed of light. Our results imply that there are more possibilities in the framework of spatially covariant gravity in light of the propagation speed of the gravitational waves.
	\end{abstract}
	
	\maketitle
%	\tableofcontents

%%%%%%
\section{Introduction}

New era of astrophysics and cosmology has arrived since the first detection of the gravitational wave (GW) event GW150914 reported by LIGO \cite{Abbott:2016blz}, and the observation of more GW events by LIGO and VIRGO \cite{Abbott:2016nmj,Abbott:2017vtc,Abbott:2017oio,Abbott:2017gyy,Abbott:2016ymx}.
With the current and the forthcoming GW experiments, including LISA \cite{Audley:2017drz}, BBO \cite{Harry:2006fi}, KAGRA \cite{Kawamura:2011zz}, ET \cite{Sathyaprakash:2012jk}, Taiji \cite{Hu:2017mde,Guo:2018npi} and Tian-Qin \cite{Luo:2015ght}, the GWs have opened a new window to explore the nature of gravity \cite{Will:2014kxa,TheLIGOScientific:2016src,Yunes:2016jcc,Cai:2017cbj}.

Among various observables concerning the gravitational waves, one that is of particular interest and of physical importance is the propagation speed of the gravitational waves. 
In practice, the propagation speed of GWs can be measured by comparing arrival times between the GWs and high-energy photons emitted at the same time of events such as the binary neutron star coalescences \cite{Nishizawa:2014zna,Nishizawa:2016kba}.
The recent observations of a binary neutron star coalescence event GW170817 \cite{TheLIGOScientific:2017qsa} and the associated gamma-ray burst event GRB170817A \cite{Monitor:2017mdv} indicates that the propagation speed of the gravitational waves coincides with the speed of light with deviations\footnote{Throughout this paper, we work in the unit with $G=c=1$.}
	\begin{equation}
	-3\times 10^{-15} \leq c_{\mathrm{T}} - 1 \leq 7\times 10^{-16} \label{cT_cons}
	\end{equation}
at the redshift $z\leq 0.009$ and with frequency of 10-100 Hz. 
The General Relativity (GR) propagates two massless tensor polarizations with the speed of light.
In gravity theories alternative to the GR, the gravitational waves propagate in a speed different from unity generally \cite{Gao:2011qe,Gao:2011vs,Saltas:2014dha,Bellini:2014fua,Gleyzes:2014rba}.
As a result, the propagation speed of the gravitational waves provides us a unique test of modified gravity theories \cite{Lombriser:2015sxa,Lombriser:2016yzn,McManus:2016kxu,Bettoni:2016mij}.

One modified gravity theory that is extensively studied in the recent years is the scalar-tensor theory, which is based on the idea of introducing an extra scalar degree of freedom in addition to the usual tensor degrees of freedom of GR.
In the theoretical aspects, recent development of the covariant scalar-tensor theory focused on introducing higher derivatives without the Ostrogradsky ghost \cite{Woodard:2015zca} (also dubbed as being ``healthy''). The representatives are the Horndeski theory \cite{Horndeski:1974wa,Deffayet:2011gz,Kobayashi:2011nu} as well as the degenerate higher-order theory \cite{Gleyzes:2014dya,Gleyzes:2014qga,Langlois:2015cwa,Motohashi:2016ftl} (see Refs. \cite{Langlois:2018dxi,Kobayashi:2019hrl} for reviews).  
After taking into account the constraint (\ref{cT_cons}), the viable Horndeski Lagrangian is \cite{Creminelli:2017sry,Sakstein:2017xjx,Ezquiaga:2017ekz,Baker:2017hug,Amendola:2017orw,Langlois:2017dyl} (see Ref. \cite{Ezquiaga:2018btd} for a review)
	\begin{equation}
	\mathcal{L}_{c_{\mathrm{T}}=1}=f(\phi) \,{}^{4}\!R+P(\phi,X)+Q(\phi,X)\square\phi,
	\end{equation}
where ${}^{4}\!R$ is the 4-dimensional Ricci scalar, $f$ is a general function of the scalar field $\phi$ only, $P$ and $Q$ are general functions of $\phi$ and $X\equiv -\frac{1}{2}(\partial\phi)^2$, and $\square \phi \equiv \nabla_{\mu}\nabla^{\mu}\phi$.
Surprisingly, the quadratic and the cubic Horndeski terms, which attracted much attention in the past decade, are completely suppressed.
See also Refs. \cite{Nunes:2018zot,Peirone:2019yjs} for recent constraints from the gravitational waves on the Horndeski theory and beyond, and Refs. \cite{Battye:2018ssx,Amendola:2018ltt,Copeland:2018yuh,Cai:2018rzd} for other modified gravity theories in which the constraint (\ref{cT_cons}) is satisfied.

An alternative approach to the scalar-tensor theory is to construct gravity theories that do not respect the full symmetry of GR.
This idea can be traced back to the effective field theory of inflation \cite{Creminelli:2006xe,Cheung:2007st} and of dark energy \cite{Creminelli:2008wc,Gubitosi:2012hu,Bloomfield:2012ff,Gleyzes:2013ooa,Bloomfield:2013efa,Gleyzes:2014rba,Gleyzes:2015pma,Gleyzes:2015rua}, as well as to the Ho\v{r}ava gravity \cite{Horava:2009uw,Blas:2009qj}. 
We may refer to such theories as spatially covariant gravity since they are metric theories respecting the spatial symmetry. 
This idea was further explored in Ref. \cite{Gao:2014soa,Gao:2014fra}, in which a general framework for the spatially covariant gravity theories was proposed.
This framework was further generalized by including a dynamical lapse function \cite{Gao:2018znj,Gao:2019lpz}, and by including an additional nondynamical scalar field \cite{Gao:2018izs}.
The virtue of the spatially covariant gravity theories is that they can be related to the scalar-tensor theories that are healthy in the unitary gauge (a gauge in which $\phi = \phi(t)$) \cite{DeFelice:2018mkq}, which are much broader than the healthy covariant scalar-tensor theories and have much more applications in cosmology.

The purpose of this work is to investigate the propagation of the gravitational waves in the general framework of spatially covariant gravity theories.
The general formulation of the propagation of the gravitational waves in modified gravity theories has been studied in Refs. \cite{Nishizawa:2017nef,Arai:2017hxj,Belgacem:2017ihm,Belgacem:2018lbp}.
Recently, the constraints on Lorentz violating gravity from the gravitational waves were investigated in Refs. \cite{Gumrukcuoglu:2017ijh,Sotiriou:2017obf,Gong:2018vbo,Mewes:2019dhj} (see also \cite{Mirshekari:2011yq}).
Comparing with the previous studies (e.g.) \cite{Fujita:2015ymn}, we shall pay special attention to two aspects which are in principle included in Ref. \cite{Gao:2014soa,Gao:2014fra} but were less studied.
\begin{itemize}
	\item Firstly, we include spatial derivatives of the extrinsic curvature (e.g., $\nabla_{k}K_{ij}$) explicitly, which were also considered in Refs. \cite{Qiao:2018dpp,Ashoorioon:2018uey,Ashoorioon:2018ocr} recently.
	Previous studies mainly focused on higher spatial derivatives of the spatial curvature $R_{ij}$, which contribute to the ``potential'' term of the tensor perturbations and result in the dispersion relation in the form $c_2k^2 + c_4k^4 +c_6k^6+\cdots$, where $k$ is the wave number.
	Higher spatial derivatives of the extrinsic curvature are also natural objects in our framework, which contribute to the ``kinetic'' terms of the tensor perturbations.
	Higher spatial derivatives of the extrinsic curvature also arise in scalar-tensor theories that are healthy in the unitary gauge \cite{Deruelle:2012xv,Crisostomi:2017ugk}.
	As we shall see, spatial derivatives of the extrinsic curvature will modify the dispersion relation of the gravitational waves in a different manner.
	\item Secondly, we introduce the parity-violating terms. 
	Including the parity-violating terms has the potential ``risk'', since generally the two polarization modes of the gravitational waves behave differently, which may conflict with the observation if we consider (\ref{cT_cons}) to hold for both polarization modes. 
	The parity-violating gravity theories and the chiral gravitational waves were extensively studied for the Chern-Simons (CS) gravity \cite{Jackiw:2003pm,Satoh:2007gn,Myung:2014jha,Bartolo:2017szm,Bartolo:2018elp,Nair:2019iur}, with Lorentz breaking \cite{Takahashi:2009wc,Myung:2009ug,Wang:2012fi,Zhu:2013fja,Cannone:2015rra}, with gauge field(s) \cite{Adshead:2013qp,Bielefeld:2014nza,Maleknejad:2016qjz,Obata:2016oym,Obata:2016xcr,Obata:2016tmo}, with chiral fermions \cite{Anber:2016yqr}, for non-Gaussianities \cite{Soda:2011am,Anber:2012du} as well as in the observational aspects \cite{Lue:1998mq,Gluscevic:2010vv,Yunes:2010yf,Smith:2016jqs,Gerbino:2016mqb,Alexander:2017jmt,Yagi:2017zhb,Thorne:2017jft,Inomata:2018rin}.
	On the other hand, it is well-known that in CS gravity the two polarization modes of GWs propagate in the speed of light \cite{Lue:1998mq,Yunes:2010yf,Alexander:2017jmt,Nishizawa:2018srh}. 
	It is thus interesting to examine if there are more general parity-violating gravity theories that have this property, i.e., satisfy the requirement $c_{\mathrm{T}} = 1$ for both polarization modes.
\end{itemize}

We will examine the viable theories under the condition $c_{\mathrm{T}} = 1$ in the broader framework of spatially covariant gravity \cite{Gao:2014soa,Gao:2014fra}.
Due to the above two aspects, the propagation speeds of the two polarization modes with helicity $s=\pm 2$ become, schematically 		
	\begin{equation}
	\left(c_{\mathrm{T}}^{(\pm2)}\right)^{2}=\frac{\mathcal{W}_{0}\pm\mathcal{W}_{1}\tilde{k}+\mathcal{W}_{2}\tilde{k}^{2}\pm\mathcal{W}_{3}\tilde{k}^{3}+\mathcal{W}_{4}\tilde{k}^{4}+\cdots}{\mathcal{G}_{0}\pm\mathcal{G}_{1}\tilde{k}+\mathcal{G}_{2}\tilde{k}^{2}\pm\mathcal{G}_{3}\tilde{k}^{3}+\mathcal{G}_{4}\tilde{k}^{4}+\cdots}, \label{cT_form}
	\end{equation}
where $\mathcal{W}_{n}, \mathcal{G}_{n}$ etc. are functions of time and $\tilde{k}\equiv k/a$ with $a$ being the scale factor.
This type of propagation speed arises in the studies of curvature perturbation of Ho\v{r}ava gravity \cite{Blas:2009qj}, and in the more general framework of spatially covariant gravity \cite{Fujita:2015ymn}. To our knowledge, for the first time in this paper we consider systematically the propagation speed in the form (\ref{cT_form}) for the gravitational waves.
In this work, we will pay special attention to the case with $c_{\mathrm{T}} = 1$, in light of the constraint (\ref{cT_cons}).
As we shall see, propagation speed in the form (\ref{cT_form}) brings us more possibilities to tune the parameters in order to have $c_{\mathrm{T}} = 1$. As a result, there is a large class of spatially covariant gravity theories in which both polarization modes of the GWs propagate with $c_{\mathrm{T}} = 1$, even in the presence of parity-violating terms.

The paper is organized as following.
In the next section, we briefly review the framework of spatially covariant gravity and show that scalar-tensor theories, including the Horndeski theory and those being healthy only in the unitary gauge, take the form of spatially covariant gravity in the unitary gauge with $\phi = t$.
In Sec. \ref{sec:propgw}, we set up our formalism by deriving the general form of the quadratic action for the gravitational waves in the cosmological background.
In Sec. \ref{sec:speed}, we focus on the propagation speed of the gravitational waves. We will examine under which conditions the two circular polarization modes of the gravitational waves could propagate in the same speed, and in particular, in the speed of light.
We identify a large class of spatially covariant gravity theories, in which both polarization modes propagate in the speed of light, even in the presence of parity-violating terms in the original action.
Section \ref{sec:con} concludes.

\emph{Notations}: Throughout this work, ${}^{4}\! R_{\mu\nu\rho\sigma}$ and ${}^{4}\! R$ stand for the 4-dimensional Riemann tensor and Ricci scalar, $R_{ij}$ and $R$ stand for the 3-dimensional Ricci tensor and scalar, respectively.

%%%%%%
\section{General framework for spatially covariant gravity theories}

\subsection{Spatially covariant theories of gravity}

Our starting point is the general action
	\begin{equation}
		S:=\int\mathrm{d}t\mathrm{d}^{3}x\,N\sqrt{h}\,\mathcal{L}\left(t,N,h_{ij},K_{ij},R_{ij},\varepsilon_{ijk},\nabla_{i}\right), \label{action}
	\end{equation}
where $N$ is the lapse function, $h_{ij}$ is the 3-dimensional spatial metric, $K_{ij}$ is the extrinsic curvature defined by
	\begin{equation}
		K_{ij} = \frac{1}{2N}\left(\dot{h}_{ij}-\pounds_{\vec{N}}h_{ij}\right),
	\end{equation}
with $\dot{h}_{ij}\equiv \partial_{t}h_{ij}$ and $\pounds_{\vec{N}}$ the Lie derivative with respect to the shift vector $N^{i}$,
$R_{ij}$ is the 3-dimensional spatial Ricci tensor, $\nabla_{i}$ is the covariant derivative compatible with $h_{ij}$.
The action (\ref{action}) generalize the framework in \cite{Gao:2014soa} by including the Levi-Civita tensor $\varepsilon_{ijk}\equiv\sqrt{h}\epsilon_{ijk}$ with $\epsilon_{123}=1$.

The theory described by (\ref{action}) is proved to propagate up to 3 dynamical degrees of freedom in the absence of $\varepsilon_{ijk}$, through a Hamiltonian analysis \cite{Gao:2014fra}. 
From the analysis it is clear that the presence of $\varepsilon_{ijk}$ does not change the constraint structure of the theory, and thus we conclude that (\ref{action}) contains at most 3 dynamical degrees of freedom.

In general, one may further extend the framework by including $\dot{N}$ in the Lagrangian through
	\begin{equation}
		F = \frac{1}{N}\left(\dot{N}-\pounds_{\vec{N}}N\right).
	\end{equation}
In this case, both the spatial metric $h_{ij}$ and the lapse function $N$ become dynamical, and there are 4 dynamical degrees of freedom (2 tensor and 2 scalar). 
As being proved in details \cite{Gao:2018znj,Gao:2019lpz}, extra conditions must be imposed in order to ensure that a single scalar degree of freedom is present.
It was shown that the resulting theory, at least up to the quadratic order in $K_{ij}$ and $F$, can be reduced to the form of Eq. (\ref{action}) through disformal transformations \cite{Gao:2019lpz}. For this reason, in the current work, we concentrate on the spatially covariant gravity without $\dot{N}$, that is, Eq. (\ref{action}).

\subsection{Comparing with the existing theories}

The advantage of the action (\ref{action}) is that it not only stands for a large class of gravity theories respecting spatial symmetries, but also provides us a unifying framework for scalar-tensor theories in the so-called unitary gauge with $\phi = \phi(t)$. In the following, we will show some examples by giving their expressions in the unitary gauge explicitly, which take the form of eq.(\ref{action}).

\subsubsection{Horndeski theory}

The expression of Horndeski Lagrangian in the unitary gauge with $\phi =t$ was firstly derived in Ref. \cite{Gleyzes:2013ooa}, which can be written as \cite{Fujita:2015ymn}
	\begin{eqnarray}
	\mathcal{L}_{\mathrm{H}}^{\mathrm{(u.g.)}} & = & a_{0}\,K-2a_{1}\,G_{ij}K^{ij}+b\left(K_{ij}K^{ij}-K^{2}\right)+\nonumber \\
	&  & +c\left(K^{3}-3KK_{ij}K^{ij}+2K_{j}^{i}K_{k}^{j}K_{i}^{k}\right)+d_{0}+d_{1}\, R,\label{LH_ug}
	\end{eqnarray}
up to the boundary terms. The six coefficients $a_{0},a_{1},b,c,d_{0},d_{1}$ are functions of $t$ and $N$, which are subject to the relations
	\begin{eqnarray}
	a_{0} & = & \frac{\partial F_{3}}{\partial N}-\frac{2}{N}\frac{\partial G_{4}}{\partial t}, \label{Hdsk_a0}\\
	2a_{1} & = & \frac{F_{5}}{N},\\
	b & = & \frac{\partial\left(NG_{4}\right)}{\partial N}+\frac{1}{2N^{2}}\frac{\partial G_{5}}{\partial t},\\
	c & = & -\frac{1}{6}\frac{\partial G_{5}}{\partial N}, \label{Hdsk_c}\\
	d_{0} & = & G_{2}+\frac{1}{N^{2}}\frac{\partial F_{3}}{\partial t}, \label{Hdsk_d0}\\
	d_{1} & = & G_{4}-\frac{1}{2N^{2}}\frac{\partial\left(G_{5}-F_{5}\right)}{\partial t}, \label{Hdsk_d1}
	\end{eqnarray}
with $G_{2},\cdots,G_{5}$ being functions of $t$ and $N$, and $F_3$ and $F_5$ being related to $G_3$ and $G_5$ through
	\begin{equation}
	\frac{\partial}{\partial N}\left(\frac{F_{3}}{N}\right)=-\frac{G_{3}}{N^{2}},\qquad\frac{\partial}{\partial N}\left(\frac{F_{5}}{N}\right)=\frac{1}{N}\frac{\partial G_{5}}{\partial N},
	\end{equation}
respectively.
It is clear that Eq. (\ref{LH_ug}) takes the form of Eq. (\ref{action}). In fact, in the unitary gauge it is possible to relax the relations Eqs. (\ref{Hdsk_a0})-(\ref{Hdsk_d1}) for the coefficients $a_{0},a_{1}$ etc..

\subsubsection{Curvature-squared terms with a scalar field}

There are 3 quadratic polynomial invariants built of the 4-dimensional Riemann tensor,
	\begin{eqnarray}
	K_{1} & = & \,{}^{4}\! R_{\mu\nu\rho\sigma}\,{}^{4}\!R^{\mu\nu\rho\sigma}\\
	K_{2} & = & \frac{1}{2}\varepsilon_{\mu\nu\alpha\beta}\,{}^{4}\! R_{\phantom{\alpha\beta}\rho\sigma}^{\alpha\beta}\,{}^{4}\!R^{\mu\nu\rho\sigma}\equiv\frac{1}{2}P,\\
	K_{3} & = & \frac{1}{4}\varepsilon_{\mu\nu\alpha\beta}\varepsilon_{\rho\sigma}^{\phantom{\rho\sigma}\lambda\tau}\,{}^{4}\! R_{\phantom{\alpha\beta}\lambda\tau}^{\alpha\beta}\,{}^{4}\! R^{\mu\nu\rho\sigma}\equiv-\mathcal{L}_{\mathrm{GB}},
	\end{eqnarray}
where $K_1$ is the so-called Kretschmann scalar, $K_{2}$ corresponds to the Chern-Pontryagin term $P$, and $K_{3}$ corresponds to the Gauss-Bonnet term $\mathcal{L}_{\mathrm{GB}}$.
$K_{1}$ by itself is less considered in the literature, since its equations of motion are of higher order and thus it is associated with the Ostrogradsky ghosts.
It is well-known that $K_2$ and $K_{3}$ are topological invariants in 4-dimension.
In order to introduce local dynamics in 4-dimension, one choice is to couple them to a scalar field in the form $f(\phi)P$ and $f(\phi)\mathcal{L}_{\mathrm{GB}}$.
For the Gauss-Bonnet term, it is well-known that $f(\phi)\mathcal{L}_{\mathrm{GB}}$ defined a well-behaved scalar-tensor theory with a single scalar degree of freedom. In fact, it has been shown that $f(\phi)\mathcal{L}_{\mathrm{GB}}$ can be recast in the form of Horndeski theory \cite{Kobayashi:2011nu}.

For the Chern-Simons (CS) gravity \cite{Lue:1998mq,Jackiw:2003pm} (see also Ref. \cite{Alexander:2009tp} for a review), the Lagrangian is 
	\begin{equation}
	\mathcal{L}_{\mathrm{CS}}:=f\left(\phi\right)P,\qquad P:=\varepsilon^{\mu\nu\rho\sigma} \,{}^{4}\! R_{\rho\sigma\alpha\beta} \,{}^{4}\! R_{\phantom{\alpha\beta}\mu\nu}^{\alpha\beta}, \label{L_CS}
	\end{equation}
where $\varepsilon^{\mu\nu\rho\sigma}$ is the Levi-Civita tensor defined by
	\begin{equation}
	\varepsilon^{\mu\nu\rho\sigma}=\frac{1}{\sqrt{-g}}\epsilon^{\mu\nu\rho\sigma},
	\end{equation}
with $\epsilon_{0123} = -\epsilon^{0123} = 1$.
In the Arnowitt-Deser-Misner (ADM) coordinates, the Chern-Pontryagin term becomes
	\begin{eqnarray}
	P & = & 16\,\varepsilon^{ijk}\Big[\frac{1}{N}\left(\dot{K}_{li}-\pounds_{\vec{N}}K_{li}\right)\nabla_{j}K_{k}^{l}- R_{i}^{l}\nabla_{j}K_{kl}\nonumber \\
	&  & \qquad-K_{i}^{l}K_{j}^{m}\nabla_{m}K_{lk}-K_{lk}K^{lm}\nabla_{i}K_{mj}-\frac{1}{N}\nabla_{i}K_{lj}\nabla_{k}\nabla^{l}N\Big].\label{Ptyg_ADM}
	\end{eqnarray}
Please note no integration-by-parts is performed in deriving (\ref{Ptyg_ADM}).
It is thus clear that due to the presence of $\dot{K}_{ij}$, generally the CS gravity (\ref{L_CS}) propagates Ostrogradsky ghosts, which is also supported by the Hamiltonian analysis \cite{Crisostomi:2017ugk}.
Nevertheless, in the unitary gauge with $\phi = \phi(t)$, $\dot{K}_{ij}$ (together with $\pounds_{\vec{N}}K_{ij}$) can be suppressed by integrations by parts. More precisely, in the unitary gauge with $\phi = t$, one can show that
	\begin{eqnarray}
	\mathcal{L}_{\mathrm{CS}}^{\mathrm{(u.g.)}} = f(t)P & \simeq & 8\varepsilon^{ijk}f\Big(K_{il}K^{lm}\nabla_{j}K_{km}+K_{i}^{l}K_{j}^{m}\nabla_{m}K_{kl}\nonumber \\
	&  & \quad-KK_{i}^{l}\nabla_{j}K_{kl}-2R_{i}^{l}\nabla_{j}K_{kl}-\frac{1}{N}\frac{\dot{f}}{f}K_{i}^{l}\nabla_{j}K_{kl}\nonumber \\
	&  & \quad-\frac{2}{N}K_{i}^{l}K_{jl}K_{km}\nabla^{m}N-\frac{2}{N}\nabla_{i}K_{jl}\nabla_{k}\nabla^{l}N\Big), \label{CS_fP_ug}
	\end{eqnarray}
up to total derivatives, where $\dot{f} \equiv \partial f(t)/\partial t$.
It is thus clear that the CS gravity in the unitary gauge takes the form of Eq. (\ref{action}) and thus propagates a single scalar degree of freedom.

The CS gravity is not the only example of scalar-tensor theories that suffer from the Ostrogradsky ghost(s) in their covariant form but are free of ghost(s) in the unitary gauge.
Another interesting term was introduced in Ref. \cite{Deruelle:2012xv}
	\begin{equation}
	\mathcal{L}_{\mathrm{DSSY}}\equiv\,{}^{4}\! C_{\mu\nu\rho\sigma}\,{}^{4}\!C_{\alpha\beta\lambda\tau}h^{\mu\alpha}h^{\nu\beta}h^{\rho\lambda}n^{\sigma}n^{\tau}, \label{CCnn}
	\end{equation}
which is quadratic in the Weyl tensor $C_{\mu\nu\rho\sigma}$, with $n_{\mu} = -N \partial{_\mu}\phi$ and $h_{\mu\nu} = g_{\mu\nu}+n_{\mu}n_{\nu}$. In the case of a timelike gradient of the scalar field, $n_{\mu}$ is normalized to be $n_{\mu}n^{\mu} = -1$ and thus $N=1/\sqrt{-(\partial\phi)^2}$.
It terms of curvature tensors, we have
	\begin{eqnarray}
	\mathcal{L}_{\mathrm{DSSY}} & = & n^{\mu}n^{\nu}\left(\,{}^{4}\! R_{\mu}^{\phantom{\mu}\alpha\beta\lambda}\,{}^{4}\!R_{\nu\alpha\beta\lambda}-\,{}^{4}\! R_{\mu}^{\alpha}\,{}^{4}\! R_{\nu\alpha}\right)\nonumber \\
	&  & +n^{\mu}n^{\nu}n^{\rho}n^{\sigma}\left(2\,{}^{4}\! R_{\mu\phantom{\alpha}\nu}^{\phantom{\mu}\alpha\phantom{\nu}\beta}\,{}^{4}\! R_{\rho\alpha\sigma\beta}-\,{}^{4}\!R_{\mu\nu}\,{}^{4}\! R_{\rho\sigma}\right).
	\end{eqnarray}
In the unitary gauge with $\phi = t$, we find
	\begin{eqnarray}
	\mathcal{L}_{\mathrm{DSSY}}^{\mathrm{(u.g.)}} & = & -\nabla_{i}K\nabla^{i}K-\nabla_{i}K^{ik}\nabla_{j}K_{k}^{j}+2\nabla^{i}K\nabla_{j}K_{i}^{j} \nonumber\\
	&  & -2\nabla_{k}K_{ij}\nabla^{j}K^{ik}+2\nabla_{k}K_{ij}\nabla^{k}K^{ij},
	\end{eqnarray}
where again no integration-by-parts is performed. It is thus clear that $\mathcal{L}_{\mathrm{DSSY}}$ in Eq. (\ref{CCnn}) describes a healthy scalar-tensor theory in the unitary gauge, which takes the form of Eq. (\ref{action}).
At this point, we emphasize that the theory is healthy only in the unitary gauge implies that they should be understood as Lorentz breaking theories.

\subsubsection{Other parity-violating theories}

The existence of Levi-Civita in the CS gravity (\ref{L_CS}) implies the parity violation.
Some exotic parity-violating gravity theories that are healthy in the unitary gauge were found in Ref. \cite{Crisostomi:2017ugk}. In the following we briefly summarize the healthy terms by reformulating the results in a more convenient form (see Appendix \ref{app:pv} for more details).

The first class of terms are quadratic in the Riemann tensor and involve only first derivative of the scalar field. 
There are three independent combinations
	\begin{eqnarray}
	\mathcal{L}_{\mathrm{A},1} & = & \varepsilon^{\mu\nu\rho\sigma}\,{}^{4}\!R_{\rho\sigma\alpha\beta}\left(\,{}^{4}\!R_{\mu\nu\phantom{\alpha}\lambda}^{\phantom{\mu\nu}\alpha}\partial^{\beta}\phi\partial^{\lambda}\phi-\frac{1}{2}\,{}^{4}\!R_{\phantom{\alpha\beta}\mu\nu}^{\alpha\beta}\partial_{\lambda}\phi\partial^{\lambda}\phi\right), \label{pv_LA1}\\
	\mathcal{L}_{\mathrm{A},2} & = & \varepsilon^{\mu\nu\rho\sigma}\,{}^{4}\!R_{\rho\sigma\alpha\beta}\left(\,{}^{4}\!R_{\phantom{\beta}\nu}^{\beta}\partial^{\alpha}\phi\partial_{\mu}\phi-\frac{1}{8}\,{}^{4}\!R_{\phantom{\alpha\beta}\mu\nu}^{\alpha\beta}\partial_{\lambda}\phi\partial^{\lambda}\phi\right), \label{pv_LA2} \\
	\mathcal{L}_{\mathrm{A},3} & = & \varepsilon^{\mu\nu\rho\sigma}\,{}^{4}\!R_{\rho\sigma\alpha\beta}\left(\,{}^{4}\!R_{\mu\lambda}^{\phantom{ag}\alpha\beta}\partial_{\nu}\phi\partial^{\lambda}\phi-\frac{1}{4}\,{}^{4}\!R_{\phantom{ef}\mu\nu}^{\alpha\beta}\partial_{\lambda}\phi\partial^{\lambda}\phi\right), \label{pv_LA3}
	\end{eqnarray}
which are healthy in the unitary gauge. 
One can show that in the unitary gauge with $\phi = t$,
	\begin{equation}
	\mathcal{L}_{\mathrm{A},1}^{\mathrm{(u.g.)}} =  -\frac{8}{N^{2}}\varepsilon_{ijk}K^{li}K^{mj}\nabla_{m}K_{l}^{\phantom{l}k}+\frac{8}{N^{2}}\varepsilon_{ijk} R^{li}\nabla^{k}K_{l}^{\phantom{l}j}, \label{LA1_gu}
	\end{equation}
and
	\begin{eqnarray}
	\mathcal{L}_{\mathrm{A},2}^{\mathrm{(u.g.)}} & = & -\frac{2}{N^{2}}\varepsilon_{ijk}K^{li}K^{mj}\nabla_{m}K_{l}^{\phantom{l}k}-\frac{2}{N^{2}}\varepsilon_{ijk}\left(K_{m}^{\phantom{l}i}K^{lm}-KK^{li}\right)\nabla^{k}K_{l}^{\phantom{l}j}\nonumber \\
	&  & +\frac{4}{N^{2}}\varepsilon_{ijk} R^{li}\nabla^{k}K_{l}^{\phantom{l}j}, \label{LA2_ug}
	\end{eqnarray}
which are clearly of the form of Eq. (\ref{action}).
$\mathcal{L}_{\mathrm{A},3}$ is trivially healthy since in the unitary gauge
	\begin{equation}
		\mathcal{L}_{\mathrm{A},3}^{\mathrm{(u.g.)}} \equiv 0. \label{LA3_ug}
	\end{equation}
We emphasize that no integration by parts is performed in deriving Eqs. (\ref{LA1_gu})-(\ref{LA3_ug}).
We may conclude that a general linear combination of $\mathcal{L}_{\mathrm{A},1}$, $\mathcal{L}_{\mathrm{A},2}$ and $\mathcal{L}_{\mathrm{A},3}$ would be healthy in the unitary gauge, although the inclusion of $\mathcal{L}_{\mathrm{A},3}$ is actually unnecessary.

The second class of terms are linear in both the Riemann tensor and the second derivative of the scalar field. In this case there is only a single term
	\begin{equation}
		\mathcal{L}_{\mathrm{B}}=\varepsilon^{\mu\nu\rho\sigma}\,{}^{4}\! R_{\rho\sigma\alpha\beta}\nabla_{\mu} \nabla^{\beta}\phi \nabla_{\nu}\phi \nabla^{\alpha}\phi.
	\end{equation}
which reduces to
	\begin{equation}
	\mathcal{L}^{\mathrm{(u.g.)}}_{\mathrm{B}}=\frac{2}{N^{3}}\varepsilon_{ijk}K^{li}\nabla^{k}K_{l}^{j},
	\end{equation}
in the unitary gauge with $\phi = t$, and is thus healthy in the unitary gauge.
	
The third class of terms are linear in the Riemann tensor and quadratic in the second derivative of the scalar field. There are three independent terms
	\begin{eqnarray}
	\mathcal{L}_{\mathrm{C},1} & = & \varepsilon^{\mu\nu\rho\sigma}\Big[\,{}^{4}\!R_{\rho\sigma\alpha\beta}\left(\nabla^{\beta}\nabla_{\nu}\phi\nabla_{\lambda}\phi\nabla^{\lambda}\phi-2\nabla^{\beta}\nabla_{\lambda}\phi\nabla_{\nu}\phi\nabla^{\lambda}\phi\right)\nonumber \\
	&  & \qquad+4\,{}^{4}\!R_{\sigma\alpha}\nabla^{\beta}\nabla_{\nu}\phi\nabla_{\rho}\phi\nabla_{\beta}\phi\Big]\nabla^{\alpha}\nabla_{\mu}\phi, \label{pv_LC1}
	\end{eqnarray}
	and
	\begin{eqnarray}
	\mathcal{L}_{\mathrm{C},2} & \equiv & \varepsilon^{\mu\nu\rho\sigma}\Big[\,{}^{4}\!R_{\rho\sigma\alpha\beta}\left(\nabla^{\lambda}\nabla_{\nu}\phi\nabla^{\beta}\phi\nabla_{\lambda}\phi-\nabla^{\beta}\nabla_{\lambda}\phi\nabla_{\nu}\phi\nabla^{\lambda}\phi\right)\nonumber \\
	&  & \qquad+2\,{}^{4}\!R_{\sigma\alpha}\nabla^{\beta}\nabla_{\nu}\phi\nabla_{\rho}\phi\nabla_{\beta}\phi\Big]\nabla^{\alpha}\nabla_{\mu}\phi, \label{pv_LC2}
	\end{eqnarray}
	and
	\begin{equation}
	\mathcal{L}_{\mathrm{C},3}\equiv\varepsilon^{\mu\nu\rho\sigma}\left(\,{}^{4}\!R_{\rho\lambda\alpha\beta}\nabla^{\lambda}\phi\nabla_{\sigma}\phi+2\,{}^{4}\!R_{\sigma\alpha}\nabla_{\rho}\phi\nabla_{\beta}\phi\right)\nabla^{\alpha}\nabla_{\mu}\phi\nabla^{\beta}\nabla_{\nu}\phi, \label{pv_LC3}
	\end{equation}
up to the quadratic order in the first derivative $\nabla_{\mu}\phi$.
In the unitary gauge with $\phi = t$, we have
	\begin{equation}
	\mathcal{L}_{\mathrm{C},1}^{\mathrm{(u.g.)}}=\frac{4}{N^{4}}\varepsilon_{ijk}\left(K^{li}K^{mj}\nabla_{m}K_{l}^{\phantom{b}k}-\frac{1}{N}K^{lj} R_{l}^{\phantom{l}k}\nabla^{i}N\right), \label{LC1_ug}
	\end{equation}
while $\mathcal{L}_{\mathrm{C},2}$ and $\mathcal{L}_{\mathrm{C},3}$ are trivial since
	\begin{equation}
	\mathcal{L}_{\mathrm{C},2}^{\mathrm{(u.g.)}}=0,\qquad\mathcal{L}_{\mathrm{C},3}^{\mathrm{(u.g.)}}=\frac{1}{2}\mathcal{L}_{1}^{\mathrm{(u.g.)}}. \label{LC2LC3_ug}
	\end{equation}
Again, no integration by parts is performed in deriving Eqs. (\ref{LC1_ug}) and (\ref{LC2LC3_ug}).
We may conclude that a general linear combination of $\mathcal{L}_{\mathrm{C},1}$, $\mathcal{L}_{\mathrm{C},2}$ and $\mathcal{L}_{\mathrm{C},3}$ is healthy in the unitary gauge, although the inclusion of $\mathcal{L}_{\mathrm{C},2}$ and $\mathcal{L}_{\mathrm{C},3}$ is actually unnecessary when working in the unitary gauge.

%%%%%
\section{Propagation of the gravitational waves} \label{sec:propgw}

In the above we have shown that scalar-tensor theories that are ghost-free in the unitary gauge typically take the form of eq.(\ref{action}).
In other words, gravity theories respecting the spatial covariance (\ref{action}) provide us a unifying framework to study modified gravity with a single scalar degree of freedom.
In this section, we investigate the propagation of linear gravitational waves within the framework of Eq.(\ref{action}) in a cosmological background.

Perturbations must respect the symmetries of the background the live on. The quadratic action for the tensor perturbations $\gamma_{ij}$ on the Friedmann-Lema\^{i}tre-Robertson-Walker (FLRW) background must take the structure:
	\begin{equation}
	S_{2}=\int\!\mathrm{d}t\mathrm{d}^{3}x\,\frac{a^{3}}{2}\left(\dot{\gamma}_{ij}\hat{\mathcal{G}}^{ij,kl}\dot{\gamma}_{kl}+\dot{\gamma}_{ij}\hat{\mathcal{F}}^{ij,kl}\gamma_{kl}-\gamma_{ij}\hat{\mathcal{W}}^{ij,kl}\gamma_{kl}\right), \label{S2_gw}
	\end{equation}
where $\hat{\mathcal{G}}^{ij,kl}$, $\hat{\mathcal{F}}^{ij,kl}$ and
$\hat{\mathcal{W}}^{ij,kl}$ are tensorial operators respecting the SO(3) symmetry of the isotropic and homogeneous spatial background. 
Throughout this work, spatial indices of perturbation quantities are raised and lowered by $\delta^{ij}$ and $\delta_{ij}$.
Note $S_{2}$ in Eq. (\ref{S2_gw}) contains only the first order time derivative of $h_{ij}$ since the original action (\ref{action}) contains only extrinsic curvature $K_{ij}$, while higher spatial derivatives are generally allowed. 
Here a hat simply reminds us that they contain spatial derivatives in general.
Without loss of generality, we assume $\hat{\mathcal{G}}^{ij,kl}$, $\hat{\mathcal{W}}^{ij,kl}$  and $\hat{\mathcal{F}}^{ij,kl}$ obey the following (anti-)symmetries under the permutations of indices:
	\begin{eqnarray}
	\hat{\mathcal{G}}^{ij,kl} & = & \hat{\mathcal{G}}^{ji,kl}=\hat{\mathcal{G}}^{ij,lk}=\hat{\mathcal{G}}^{kl,ij}, \label{calG_sym}\\
	\hat{\mathcal{W}}^{ij,kl} & = & \hat{\mathcal{W}}^{ji,kl}=\hat{\mathcal{W}}^{ij,lk}=\hat{\mathcal{W}}^{kl,ij}, \label{calW_sym}
	\end{eqnarray}
and
	\begin{equation}
	\hat{\mathcal{F}}^{ij,kl}=\hat{\mathcal{F}}^{ji,kl}=\hat{\mathcal{F}}^{ij,lk}=-\hat{\mathcal{F}}^{kl,ij}. \label{calF_as}
	\end{equation}
The last equality is because the symmetric part of $\hat{\mathcal{F}}^{ij,kl}$, if exists, can always be reduced by integrations by parts. 

Within the framework of Eq. (\ref{action}), $\hat{\mathcal{G}}^{ij,kl}$, $\hat{\mathcal{F}}^{ij,kl}$ and $\hat{\mathcal{W}}^{ij,kl}$ must be built of the background quantities. In our case, these are
 	\begin{equation}
	 	\delta^{ij},\qquad \epsilon^{ijk},\qquad \text{with time-dependent functions},
 	\end{equation}
together with spatial derivatives.
There are only two basic operators that have non-vanishing contributions to the quadratic action of tensor perturbations:
	\begin{equation}
	S^{ij,kl}:=\frac{1}{2}\left(\delta^{ik}\delta^{jl}+\delta^{il}\delta^{jk}\right),
	\end{equation}
and $A^{ij,kl,m}\partial_{m}$ with
	\begin{equation}
	A^{ij,kl,m}:=\frac{1}{4}\left(\delta^{ik}\epsilon^{jlm}+\delta^{il}\epsilon^{jkm}+\delta^{jk}\epsilon^{ilm}+\delta^{jl}\epsilon^{ikm}\right).
	\end{equation}
Both $S^{ij,kl}$ and $A^{ij,kl,m}\partial_{m}$ satisfy the relations in (\ref{calG_sym}) and (\ref{calW_sym}).	
It is not possible, however, to build an operator of $S^{ij,kl}$ and $A^{ij,kl,m}$ with spatial derivatives satisfying all the relations (in particular, the last equality) in Eq. (\ref{calF_as}), which implies that 
	\begin{equation}
		\hat{\mathcal{F}}^{ij,kl} \equiv 0,
	\end{equation}
in our case.

We thus conclude that in our framework of spatially covariant gravity described by eq.(\ref{action}), the quadratic action for the gravitational waves in a FLRW background takes the form
	\begin{equation}
	S_{2}=\int\mathrm{d}t\mathrm{d}^{3}x\,\frac{a^{3}}{2}\left(\dot{\gamma}_{ij}\hat{\mathcal{G}}^{ij,kl}\dot{\gamma}_{kl}-\gamma_{ij}\hat{\mathcal{W}}^{ij,kl}\gamma_{kl}\right), \label{S2T_gen}
	\end{equation}
where $\hat{\mathcal{G}}^{ij,kl}$ and $\hat{\mathcal{W}}^{ij,kl}$ are built of $S^{ij,kl}$, $A^{ij,kl,m}\partial_{m}$ and spatial derivatives.
In this paper, we further assume that $\hat{\mathcal{G}}^{ij,kl}$ and $\hat{\mathcal{W}}^{ij,kl}$ can be expanded as series of spatial derivatives, which take the general form
	\begin{equation}
	\hat{\mathcal{G}}^{ij,kl}\left(t,\bm{\partial}\right)=\sum_{n=0}\left(\mathcal{G}_{2n}S^{ij,kl}-\frac{1}{a}\mathcal{G}_{2n+1}A^{ij,kl,m}\partial_{m}\right)\frac{\left(-\Delta\right)^{n}}{a^{2n}}, \label{hatcalG}
	\end{equation}
and
	\begin{equation}
	\hat{\mathcal{W}}^{ij,kl}\left(t,\bm{\partial}\right)=\sum_{n=0}\left(\mathcal{W}_{2n}S^{ij,kl}-\frac{1}{a}\mathcal{W}_{2n+1}A^{ij,kl,m}\partial_{m}\right)\frac{\left(-\Delta\right)^{n+1}}{a^{2n+2}}, \label{hatcalW}
	\end{equation}
with $\Delta = \delta^{ij}\partial_{i}\partial_{j}$.
Here $\mathcal{G}_{n}$ and $\mathcal{W}_{n}$ are general functions of time, of which the explicit expression depend on the concrete models.
From the above, $\mathcal{G}_{n}$ and $\mathcal{W}_{n}$ with $n=0,2,4,6,\cdots$ are parity-preserving terms, and $\mathcal{G}_{n}$ and $\mathcal{W}_{n}$ with $n=1,3,5,7,\cdots$ are parity-violating terms.

In the Fourier space, the quadratic action (\ref{S2T_gen}) becomes
	\begin{equation}
	S_{2}=\int\mathrm{d}t\frac{\mathrm{d}^{3}k}{\left(2\pi\right)^{3}}\,\frac{a^{3}}{2}\left(\dot{\gamma}_{ij}\left(t,\bm{k}\right)\hat{\mathcal{G}}^{ij,kl}\left(t,-i\bm{k}\right)\dot{\gamma}_{kl}\left(t,-\bm{k}\right)-\gamma_{ij}\left(t,\bm{k}\right)\hat{\mathcal{W}}^{ij,kl}\left(t,-i\bm{k}\right)\gamma_{kl}\left(t,-\bm{k}\right)\right),\label{S2_gen_k}
	\end{equation}
with 
	\begin{equation}
	\hat{\mathcal{G}}^{ij,kl}\left(t,-i\bm{k}\right) = \sum_{n=0}\left(\mathcal{G}_{2n}\left(t\right)S^{ij,kl}-\mathcal{G}_{2n+1}\left(t\right)A^{ij,kl,m}\frac{\left(-i\right)\bm{k}_{m}}{a}\right)\frac{k^{2n}}{a^{2n}},
	\end{equation}
and
	\begin{equation}
	\hat{\mathcal{W}}^{ij,kl}\left(t,-i\bm{k}\right) = \sum_{n=0}\left(\mathcal{W}_{2n}\left(t\right)S^{ij,kl}-\mathcal{W}_{2n+1}\left(t\right)A^{ij,kl,m}\frac{\left(-i\right)\bm{k}_{m}}{a}\right)\frac{k^{2n+2}}{a^{2n+2}}.
	\end{equation}

We decompose the tensor perturbation $\gamma_{ij}$ into the polarization modes:
	\begin{equation}
	\gamma_{ij}(t,\bm{k})=\sum_{s=\pm2}e_{ij}^{(s)}(\hat{\bm{k}})\gamma^{(s)}(t,\bm{k}),\label{polar_dec}
	\end{equation}
where $\hat{\bm{k}} \equiv \bm{k}/|\bm{k}|$, $e_{ij}^{\left(s\right)}(\hat{\bm{k}})$ are the circular polarization tensors with the helicity states $s=\pm2$, satisfying the traceless and transverse conditions
	\begin{equation}
	\delta^{ij}e_{ij}^{(s)}(\hat{\bm{k}}) = k^{i}e_{ij}^{(s)}(\hat{\bm{k}})=0.\label{eij_TT}
	\end{equation}
We follow the convention in Ref. \cite{Gao:2011vs} and choose the phase of $e_{ij}^{(s)}(\hat{\bm{k}})$ such that 
	\begin{equation}
	e_{ij}^{(s)\ast}(\hat{\bm{k}}) = e_{ij}^{(-s)}(\hat{\bm{k}})=e_{ij}^{(s)}(-\hat{\bm{k}}),\label{eij_pr}
	\end{equation}
where an asterisk denotes the complex conjugate.
The two polarization tensors are normalized to be
	\begin{equation}
	e_{ij}^{(s)}(\hat{\bm{k}})e^{(-s')ij}(\hat{\bm{k}})=\delta^{ss'}.\label{ee_nom}
	\end{equation}
Make use of the relation \cite{Alexander:2004wk,Satoh:2007gn,Bartolo:2017szm}
	\begin{equation}
	i\hat{\bm{k}}^{l}\epsilon_{lij}e_{m}^{(s)i}(\hat{\bm{k}})e^{(s')jm}(-\hat{\bm{k}})=\frac{s}{2}\delta^{ss'},\label{ee_epsi}
	\end{equation}
after some manipulations, the quadratic action for the polarization modes is
	\begin{eqnarray}
	S_{2} & = & \int\!\mathrm{d}\tau\frac{\mathrm{d}^{3}k}{\left(2\pi\right)^{3}}\,\frac{a^{2}}{2}\sum_{s=\pm2}\mathcal{G}^{(s)}\left(\tau,k\right)\Big(\partial_{\tau}\gamma^{(s)}\left(\tau,\bm{k}\right)\partial_{\tau}\gamma^{(s)}\left(\tau,-\bm{k}\right)\nonumber \\
	&  & \hspace{10em}-k^{2}\frac{\mathcal{W}^{(s)}\left(\tau,k\right)}{\mathcal{G}^{(s)}\left(\tau,k\right)}\gamma^{(s)}\left(\tau,\bm{k}\right)\gamma^{(s)}\left(\tau,-\bm{k}\right)\Big).\label{S2_gw_gen}
	\end{eqnarray}
where $\tau$ is the conformal time defined by $\mathrm{d}t = a \mathrm{d}\tau$, and
	\begin{eqnarray}
	\mathcal{G}^{(s)}\left(\tau,k\right) & := & \sum_{n=0}\mathcal{G}_{n}\left(\tau\right)\left(\frac{s}{2}\frac{k}{a}\right)^{n}, \label{calG_s}\\
	\mathcal{W}^{(s)}\left(\tau,k\right) & := & \sum_{n=0}\mathcal{W}_{n}\left(\tau\right)\left(\frac{s}{2}\frac{k}{a}\right)^{n}. \label{calW_s}
	\end{eqnarray}
From (\ref{S2_gw_gen}), the two circular polarization modes of the gravitational waves are decoupled, even if the parity-violating terms are present. Eq. (\ref{S2_gw_gen}) (together with Eqs.(\ref{calG_s}) and (\ref{calW_s})) is one of the main results in this paper.

The equations of motion for the polarization modes are thus
	\begin{equation}
	\partial_{\tau}^{2}\gamma^{(s)}\left(\tau,\bm{k}\right)+\mathcal{H}\left(2+\nu^{(s)}\right)\partial_{\tau}\gamma^{(s)}\left(\tau,\bm{k}\right)+\left(c_{\mathrm{T}}^{(s)}\right)^{2}k^{2}\gamma^{(s)}\left(\tau,\bm{k}\right)=0,\qquad s=\pm2.\label{eom_gammas}
	\end{equation}
where $\mathcal{H}$ is the comoving Hubble parameter defined by $\mathcal{H} = \partial_{\tau}\ln a$,
	\begin{eqnarray}
	\nu^{(s)}\left(\tau,k\right) & = & \frac{1}{\mathcal{H}}\frac{\partial_{\tau}\mathcal{G}^{(s)}\left(\tau,k\right)}{\mathcal{G}^{(s)}\left(\tau,k\right)}, \label{nu_s}\\
	\left(c_{\mathrm{T}}^{(s)}\left(\tau,k\right)\right)^{2} & = & \frac{\mathcal{W}^{(s)}\left(\tau,k\right)}{\mathcal{G}^{(s)}\left(\tau,k\right)}. \label{cT_s}
	\end{eqnarray}
Eq. (\ref{eom_gammas}) can be compared with the relevant equations in Ref. \cite{Saltas:2014dha} (see also Refs. \cite{Nishizawa:2017nef,Nishizawa:2018srh}). The parameter $\nu^{(s)}$ is identified to be the Planck mass running rate, which modifies the amplitude of the gravitational waves and is related to the strength of the gravity. 
The effect of a running Planck mass on the GWs was recently discussed in Ref. \cite{Lagos:2019kds}. The parameter $c_{\mathrm{T}}^{(s)}$ is identified to be the propagation speed (phase velocity) of the gravitational waves. 
In the case of GR, $\nu^{(s)} = 0$ and $c_{\mathrm{T}}^{(s)} = 1$.
In a general modified theory of gravity, both $\nu^{(s)}\left(\tau,k\right)$ and $c_{\mathrm{T}}^{(s)}\left(\tau,k\right)$ depend on the time $\tau$, the wave number $k$ and the helicity $s$.

%%%%%%
\section{Propagation speed of the gravitational waves} \label{sec:speed}

Massless fields must propagate in the same speed as that of the light, as demanded by the Einstein equivalence principle (EEP). 
However, EEP says nothing about the propagation speed of gravitational waves, which may vary from theory to theory.
This has been known for theories respecting general covariance, e.g., Horndeski theory \cite{Gao:2011qe}, in which the gravitational waves obey a linear dispersion relation.
For Lorentz breaking gravity theories, the gravitational waves may even propagate with nonlinear dispersion relations.

We will pay special attention to the propagation speed of gravitational waves in our framework. 
For concreteness, we consider $\mathcal{G}^{(s)}$ and $\mathcal{W}^{(s)}$ up to $k^{4}$, which corresponds to the case where the original action (\ref{action}) is up to the sixth order in derivatives. 
From Eqs. (\ref{calG_s})-(\ref{calW_s}), the propagation speeds of the polarization modes are thus
	\begin{equation}
	\left(c_{\mathrm{T}}^{(s)}\right)^{2}=\frac{\mathcal{W}_{0}\left(t\right)+\mathcal{W}_{1}\left(t\right)\frac{s}{2}\frac{k}{a}+\mathcal{W}_{2}\left(t\right)\frac{k^{2}}{a^{2}}+\mathcal{W}_{3}\left(t\right)\frac{s}{2}\frac{k^{3}}{a^{3}}+\mathcal{W}_{4}\left(t\right)\frac{k^{4}}{a^{4}}}{\mathcal{G}_{0}\left(t\right)+\mathcal{G}_{1}\left(t\right)\frac{s}{2}\frac{k}{a}+\mathcal{G}_{2}\left(t\right)\frac{k^{2}}{a^{2}}+\mathcal{G}_{3}\left(t\right)\frac{s}{2}\frac{k^{3}}{a^{3}}+\mathcal{G}_{4}\left(t\right)\frac{k^{4}}{a^{3}}}. \label{cT2_k4}
	\end{equation}
Comparing with the propagation speed that arises in usual case (e.g., in the Horndeski theory), the propagation speed in the form of Eq. (\ref{cT2_k4}) has at least two distinctive features.
	\begin{itemize}
		\item Generally, the two circular polarization modes propagate in different speeds, i.e., $c_{\mathrm{T}}^{(+2)} \neq c_{\mathrm{T}}^{(-2)}$, due to the presence of parity-violating terms. However, it is possible that the coefficients in the original action are turned such that the two polarization modes propagate in the same speed in the presence of parity-violating terms. This has been known for the case of CS gravity \cite{Lue:1998mq,Yunes:2010yf,Alexander:2017jmt,Nishizawa:2018srh}. 
		\item There are $k$-dependence in the denominator of $\left(c_{\mathrm{T}}^{(s)}\right)^{2}$. The $k$-dependence in $\mathcal{G}^{(s)}$ arises due to spatial derivative terms of the extrinsic curvature $K_{ij}$ in the original action. This is different from usual Lorentz-breaking theories such as Ho\v{r}ava gravity, where the propagation speed is a polynomial in $k$, which corresponds to $\mathcal{G}_{n} = 0$ for $n>0$. A simpler version of Eq. (\ref{cT2_k4}) arises in Ref. \cite{Nishizawa:2018srh} (see also Ref. \cite{Cai:2016ihp})in the study of a variation of the CS gravity.
	\end{itemize}
These introduce new features of the propagation of the gravitational waves. Especially, this makes the two polarization modes propagate in the same speed possible, even in the presence of parity-violating terms in the original action. In fact, as we shall see in Sec.\ref{sec:cTcl}, the CS gravity is not the only parity-violating gravity theory in which the two polarization modes of the gravitational waves propagate in the speed of light. There are more general parity-violating gravity theories have this property.

Before proceeding, let us compare our result (\ref{cT2_k4}) with the propagation speed of gravitational waves in parity-violating gravity theories studied before.
If in $\mathcal{G}^{(s)}$ and $\mathcal{W}^{(s)}$ only the parity-violating terms with the lowest order in spatial derivatives are present, Eq. (\ref{cT2_k4}) reduces to
	\begin{equation}
	\left(c_{\mathrm{T}}^{(s)}\right)^{2}=\frac{\mathcal{W}_{0}\left(t\right)+\mathcal{W}_{1}\left(t\right)\frac{s}{2}\frac{k}{a}}{\mathcal{G}_{0}\left(t\right)+\mathcal{G}_{1}\left(t\right)\frac{s}{2}\frac{k}{a}}.
	\end{equation}
In the case of CS gravity 
	\begin{equation}
		\mathcal{G}_{0} = \mathcal{W}_0 = 1,\qquad \mathcal{G}_{1} = \mathcal{W}_{1},
	\end{equation} 
which yields $c_{\mathrm{T}}^{(s)} \equiv 1$ in CS gravity.
In a more general setting with $\mathcal{G}_{0} = \mathcal{W}_0 = 1$ but $ \mathcal{G}_{1} \neq \mathcal{W}_{1}$, one get the conclusion that if one polarization mode is enhanced/superluminal, the other is suppressed/subluminal \cite{Nishizawa:2018srh}.
In general, this is not the case for the propagation speeds given in Eq. (\ref{cT2_k4}).

\subsection{On the Case of $c_{\mathrm{T}}^{(+2)}=c_{\mathrm{T}}^{(-2)}$}

As mentioned in the above, there is room of parameters such that the two polarization modes propagate with the same speed, despite of the presence of parity-violating terms in the action.
Mathematically, this is because the $s$-dependence of $c_{\mathrm{T}}^{(s)}$ can be balanced between $\mathcal{G}^{(s)}$ and $\mathcal{W}^{(s)}$. In particular, this can be achieved only if $\mathcal{G}^{(s)}$ itself has functional dependence on $s$ and $k$, which is due to the presence of spatial derivatives of the extrinsic curvature $K_{ij}$ in our framework.

For consistency, we assume $\mathcal{G}_{0}\neq0$ and $\mathcal{W}_{0}\neq0$ (otherwise the case of GR cannot be recovered).
Requiring that the two polarization modes propagate with the same speed, i.e.,
	\begin{equation}
	\left(c_{\mathrm{T}}^{(+2)}\right)^{2}=\left(c_{\mathrm{T}}^{(-2)}\right)^{2},
	\end{equation}
yields 4 constraints among the 8 coefficients $\mathcal{G}_{1},\cdots,\mathcal{G}_{4}$
and $\mathcal{W}_{1},\cdots,\mathcal{W}_{4}$:
	\begin{eqnarray}
	2\mathcal{G}_{0}\mathcal{W}_{1}-2\mathcal{G}_{1}\mathcal{W}_{0} & = & 0,\\
	-2\mathcal{G}_{3}\mathcal{W}_{0}+2\mathcal{G}_{2}\mathcal{W}_{1}-2\mathcal{G}_{1}\mathcal{W}_{2}+2\mathcal{G}_{0}\mathcal{W}_{3} & = & 0,\\
	2\mathcal{G}_{4}\mathcal{W}_{1}-2\mathcal{G}_{3}\mathcal{W}_{2}+2\mathcal{G}_{2}\mathcal{W}_{3}-2\mathcal{G}_{1}\mathcal{W}_{4} & = & 0,\\
	2\mathcal{G}_{4}\mathcal{W}_{3}-2\mathcal{G}_{3}\mathcal{W}_{4} & = & 0.
	\end{eqnarray}
There are 3 branches of solutions.
\begin{itemize}
	\item Case-1: The first branch of solutions is
	\begin{equation}
	\mathcal{G}_{1}=0,\quad\mathcal{G}_{3}=0,\quad\mathcal{W}_{1}=0,\quad\mathcal{W}_{3}=0.
	\end{equation}
	This case is trivial, since there is no parity-violating term at all. The two polarization modes propagate with the same speed
	\begin{equation}
	\left(c_{\mathrm{T}}^{(s)}\right)^{2}=\frac{\mathcal{W}_{0}+\mathcal{W}_{2}\frac{k^{2}}{a^{2}}+\mathcal{W}_{4}\frac{k^{4}}{a^{4}}}{\mathcal{G}_{0}+\mathcal{G}_{2}\frac{k^{2}}{a^{2}}+\mathcal{G}_{4}\frac{k^{4}}{a^{4}}}.
	\end{equation}
	Note the spatially covariant gravity considered in Refs. \cite{Gao:2014soa,Fujita:2015ymn} (including the Ho\v{r}ava gravity \cite{Horava:2009uw}) belongs to this case, where $\mathcal{G}_{2} = \mathcal{G}_{4} = 0$.
	\item Case-2:
	In the case with $\mathcal{G}_{1}\neq 0$, we get another branch of solutions
	\begin{eqnarray}
	\mathcal{G}_{4} & = & \frac{\mathcal{G}_{1}\mathcal{G}_{2}-\mathcal{G}_{0}\mathcal{G}_{3}}{\mathcal{G}_{1}^{2}}\mathcal{G}_{3},\\
	\mathcal{W}_{1} & = & \frac{\mathcal{G}_{1}\mathcal{W}_{0}}{\mathcal{G}_{0}},\\
	\mathcal{W}_{2} & = & \frac{\mathcal{G}_{2}\mathcal{W}_{0}}{\mathcal{G}_{0}}-\frac{\mathcal{G}_{3}\mathcal{W}_{0}}{\mathcal{G}_{1}}+\frac{\mathcal{G}_{0}\mathcal{W}_{3}}{\mathcal{G}_{1}},\\
	\mathcal{W}_{4} & = & \frac{\mathcal{G}_{1}\mathcal{G}_{2}-\mathcal{G}_{0}\mathcal{G}_{3}}{\mathcal{G}_{1}^{2}}\mathcal{W}_{3}.
	\end{eqnarray}
	The propagation speed is given by
	\begin{equation}
	\left(c_{\mathrm{T}}^{(s)}\right)^{2}=\frac{\mathcal{G}_{1}\frac{\mathcal{W}_{0}}{\mathcal{G}_{0}}+\mathcal{W}_{3}\frac{k^{2}}{a^{2}}}{\mathcal{G}_{1}+\mathcal{G}_{3}\frac{k^{2}}{a^{2}}}.
	\end{equation}
	In this case, since $\mathcal{G}_{1}, \mathcal{W}_{1} \neq 0$, the theory contains parity-violating terms. Nevertheless, the parity-violating effects do not show up in the propagation speed of the polarization modes. 
	\item Case-3: We also have a special case with
	\begin{equation}
	\mathcal{W}_{1}=\frac{\mathcal{G}_{1}\mathcal{W}_{0}}{\mathcal{G}_{0}},\quad\mathcal{W}_{2}=\frac{\mathcal{G}_{2}\mathcal{W}_{0}}{\mathcal{G}_{0}},\quad\mathcal{W}_{3}=\frac{\mathcal{G}_{3}\mathcal{W}_{0}}{\mathcal{G}_{0}},\quad\mathcal{W}_{4}=\frac{\mathcal{G}_{4}\mathcal{W}_{0}}{\mathcal{G}_{0}}.
	\end{equation}
	In this case the propagation speed is simply
	\begin{equation}
	\left(c_{\mathrm{T}}^{(s)}\right)^{2}=\frac{\mathcal{W}_{0}}{\mathcal{G}_{0}},
	\end{equation}
	in which the $k$-dependence of $c_{\mathrm{T}}^{(s)}$ completely drops out. We emphasize that ``Case-2'' does not include  ``Case-3'' as a special case.
\end{itemize}
In ``case-1'', there is no parity-violating terms by construction, and thus both polarization modes also have the same amplitude when being quantized, although the dispersion relation is highly nonlinear. This can be seen also from Eq. (\ref{nu_s}), which implies $\nu^{(2)} = \nu^{(-2)}$ in ``Case-1''. 
On the other hand, in ``case-2'' and ``case-3'', although the two polarization modes propagate in the same speed, they have different amplitudes since generally $\nu^{(2)} \neq \nu^{(-2)}$  due to the presence of parity-violating terms $\mathcal{G}_{1}$ and $\mathcal{G}_3$.

\subsection{Models with $c_{\mathrm{T}}^{(+2)}=c_{\mathrm{T}}^{(-2)}=1$} \label{sec:cTcl}

The detection of GW170817 \cite{TheLIGOScientific:2017qsa} and GRB170817A \cite{Monitor:2017mdv} indicates that the propagation speed of the gravitational waves coincides with the speed of light with tiny deviations (\ref{cT_cons}).
Limit of the same order has already reported in the gravitational Cherenkov effect \cite{Moore:2001bv}.
Although the physics of GW170817 may be different from that in the primordial universe, it has already been used to restrict the structure of scalar-tensor theories \cite{Creminelli:2017sry,Sakstein:2017xjx,Ezquiaga:2017ekz,Baker:2017hug,Langlois:2017dyl}. 
Within our framework, this corresponds to a special case of ``case-3'' in the above, which implies
	\begin{equation}
	\mathcal{W}_{n}=\mathcal{G}_{n},\qquad n=0,1,2,3,4. \label{WnGn}
	\end{equation}

In the following, we investigate a concrete model of which the Lagrangian is a polynomial built of the extrinsic curvature $K_{ij}$ and intrinsic curvature $R_{ij}$ as well as their spatial derivatives. We classify each monomial according to the orders of time and spatial derivatives of $h_{ij}$. Note $K_{ij}$ contains the first order time derivative of $h_{ij}$, $R_{ij}$ contains up to the second order in spatial derivatives of $h_{ij}$.  In Tab. \ref{tab:terms}, we list all the possible terms up to the fourth order in derivatives of $h_{ij}$.
	\begin{table}[h]
	\begin{centering}
		\begin{tabular}{|c|c|>{\raggedright}p{11cm}|}
			\hline 
			$d$ & $\left(d_{\mathrm{t}},d_{\mathrm{s}}\right)$ & operators\tabularnewline
			\hline 
			0 & $\left(0,0\right)$ & 1\tabularnewline
			\hline 
			\multirow{2}{*}{1} & $\left(1,0\right)$ & $K$\tabularnewline
			\cline{2-3} \cline{3-3} 
			& $\left(0,1\right)$ & -\tabularnewline
			\hline 
			\multirow{3}{*}{2} & $\left(2,0\right)$ & $K_{ij}K^{ij},\qquad K^{2}$\tabularnewline
			\cline{2-3} \cline{3-3} 
			& $\left(1,1\right)$ & -\tabularnewline
			\cline{2-3} \cline{3-3} 
			& $\left(0,2\right)$ & $R$\tabularnewline
			\hline 
			\multirow{4}{*}{3} & $\left(3,0\right)$ & $K_{ij}K^{jk}K_{k}^{i},\qquad K_{ij}K^{ij}K,\qquad K^{3}$\tabularnewline
			\cline{2-3} \cline{3-3} 
			& $\left(2,1\right)$ & $\varepsilon_{ijk}K_{l}^{i}\nabla^{j}K^{kl}$\tabularnewline
			\cline{2-3} \cline{3-3} 
			& $\left(1,2\right)$ & $\nabla^{i}\nabla^{j}K_{ij},\qquad\nabla^{2}K,\qquad R^{ij}K_{ij},\qquad RK$\tabularnewline
			\cline{2-3} \cline{3-3} 
			& $\left(0,3\right)$ & -\tabularnewline
			\hline 
			\multirow{5}{*}{4} & $\left(4,0\right)$ & $K_{ij}K^{jk}K_{k}^{i}K,\quad\left(K_{ij}K^{ij}\right)^{2},\quad K_{ij}K^{ij}K^{2},\quad K^{4}$\tabularnewline
			\cline{2-3} \cline{3-3} 
			& $\left(3,1\right)$ & $\varepsilon_{ijk}\nabla_{m}K_{n}^{i}K^{jm}K^{kn},\qquad\varepsilon_{ijk}\nabla^{i}K_{m}^{j}K_{n}^{k}K^{mn},\qquad\varepsilon_{ijk}\nabla^{i}K_{l}^{j}K^{kl}K$\tabularnewline
			\cline{2-3} \cline{3-3} 
			& $\left(2,2\right)$ & $\nabla_{k}K_{ij}\nabla^{k}K^{ij},\qquad\nabla_{i}K_{jk}\nabla^{k}K^{ij},\qquad\nabla_{i}K^{ij}\nabla_{k}K_{j}^{k},\qquad\nabla_{i}K^{ij}\nabla_{j}K,$
			$\nabla_{i}K\nabla^{i}K,\qquad R_{ij}K_{k}^{i}K^{jk},\qquad RK_{ij}K^{ij},\qquad R_{ij}K^{ij}K,\qquad K^{2}$\tabularnewline
			\cline{2-3} \cline{3-3} 
			& $\left(1,3\right)$ & $\varepsilon_{ijk}R^{il}\nabla^{j}K_{l}^{k},\qquad\varepsilon_{ijk}\nabla^{i}R_{l}^{j}K^{kl}$\tabularnewline
			\cline{2-3} \cline{3-3} 
			& $\left(0,4\right)$ & $\nabla^{i}\nabla^{j}R_{ij},\qquad\nabla^{2}R,\qquad R_{ij}R^{ij},\qquad R^{2}$\tabularnewline
			\hline 
		\end{tabular}
		\par\end{centering}
	\caption{All the possible monomials built of $K_{ij}$, $R_{ij}$ and their spatial derivatives, up to the fourth order in derivatives.}
	\label{tab:terms}
\end{table}
In Tab.\ref{tab:terms}, $d_{\mathrm{t}}$ and $d_{\mathrm{s}}$ are the numbers of time derivative and spatial derivative, respectively. 
We emphasize that not all the terms in the above table (e.g., $K^2$, $K^3$, $\nabla^{i}\nabla^{j}K_{ij}$ etc.) contribute to the quadratic action of the gravitational waves. There are 35 individual terms in the above table, while only 21 terms contribute to the propagation of linear gravitational waves. Second, we do not list terms involving spatial derivatives of the lapse function $N$, since which do not contribute to the quadratic action of the gravitational waves in a cosmological background.

Our starting point is the action
	\begin{equation}
	S=\int\!\mathrm{d}t\mathrm{d}^{3}x\,N\sqrt{h}\left(L^{(0)}+L^{(1)}+L^{(2)}+L^{(3)}+L^{(4)}\right),\label{model}
	\end{equation}
where $L^{(d)}$ stands for the linear combinations of terms in the above table satisfying $d_{\mathrm{t}}+d_{\mathrm{s}}=d$, such as
	\begin{eqnarray}
	L^{(0)} & = & c_{1}^{(0,0)},\label{L0}\\
	L^{(1)} & = & c_{1}^{(1,0)}K,\label{L1}
	\end{eqnarray}
and
	\begin{equation}
	L^{(2)}=c_{1}^{(2,0)}K_{ij}K^{ij}+c_{2}^{(2,0)}K^{2}+c_{1}^{(0,2)}R,\label{L2}
	\end{equation}
etc.
All the coefficients $c_{1}^{(1,0)}, c_{1}^{(2,0)}$ etc. are functions
of $t$ and $N$, e.g.,
	\begin{equation}
	c_{1}^{(1,0)}=c_{1}^{(1,0)}\left(t,N\right).
	\end{equation}
Note generally the coefficients may also depend on spatial derivatives of lapse function $N$, and there are terms involving spatial derivatives of $N$ which we do not include in Eq. (\ref{model}). Terms involving spatial derivatives of $N$ do not contribute to the linear gravitational waves in the FLRW background, although they may be considered when analysing the background evolution and scalar perturbations.

We will study the linear gravitational waves of the action Eq. (\ref{model}) around the FLRW background.
To this end, we consider the perturbed metric
	\begin{equation}
	\mathrm{d}s^{2}\equiv-\mathrm{d}t^{2}+a^{2}\mathfrak{g}_{ij}\mathrm{d}x^{i}\mathrm{d}x^{j},\label{metric_ADM}
	\end{equation}
with $a=a\left(t\right)$ being the scale-factor.
At the background level $\bar{\mathfrak{g}}_{ij} = \delta_{ij}$.
It is proved convenient to define the perturbation of $\mathfrak{g}_{ij}$
in the ``exponential'' manner:
	\begin{eqnarray}
	\mathfrak{g}_{ij} & := & \delta_{ik}\left(e^{\bm{\gamma}}\right)_{\phantom{k}j}^{k} \nonumber\\
	& = & \delta_{ij}+\gamma_{ij}+\frac{1}{2}\gamma_{ik}\gamma_{\phantom{k}j}^{k}+\cdots, \label{gij_eH}
	\end{eqnarray}
where $\gamma^{i}_{j}$ is the tensor perturbation satisfying $\partial_{i}\gamma^{i}_{\phantom{i}j} = 0$ and $\gamma^{i}_{\phantom{i}i} = 0$, and we define
	\begin{equation}
	\gamma_{ij}:=\delta_{ik}\gamma_{\phantom{k}j}^{k}.
	\end{equation}
The advantage of defining $\mathfrak{g}_{ij}$ in the exponential
manner is that $\det\mathfrak{g}_{ij}\equiv1$ (in the presence of
tensor modes only), which is unperturbed.
With Eqs. (\ref{metric_ADM}) and (\ref{gij_eH}), we consider only the tensor modes, which is justified by the fact that the scalar, vector and tensor perturbations are decoupled at the linear order in the FLRW background.

After some manipulations, the contribution of the action Eq. (\ref{model}) to the quadratic action for the tensor modes takes the form of Eq. (\ref{S2T_gen}), i.e.,
	\begin{eqnarray}
	S_{2} & = & \int\mathrm{d}t\mathrm{d}^{3}x\,\frac{a^{3}}{2}\Big(\mathcal{G}_{0}\left(t\right)\dot{\gamma}_{ij}\dot{\gamma}^{ij}+\mathcal{G}_{1}\left(t\right)\epsilon^{ijk}\dot{\gamma}_{li}\frac{1}{a}\partial_{j}\dot{\gamma}_{k}^{l}-\mathcal{G}_{2}\left(t\right)\dot{\gamma}_{ij}\frac{\Delta}{a^{2}}\dot{\gamma}^{ij} \nonumber\\
	&  & \qquad+\mathcal{W}_{0}\left(t\right)\gamma_{ij}\frac{\Delta}{a^{2}}\gamma^{ij}+\mathcal{W}_{1}\left(t\right)\epsilon^{ijk}\gamma_{li}\frac{1}{a}\frac{\Delta}{a^{2}}\partial_{j}\gamma_{k}^{l}-\mathcal{W}_{2}\left(t\right)\gamma_{ij}\frac{\Delta^{2}}{a^{4}}\gamma^{ij}\Big), \label{S2T_model}
	\end{eqnarray}
where $\mathcal{G}_{n}$ and $\mathcal{W}_{n}$ are given by
	\begin{equation}
	\mathcal{G}_{0}\left(t\right)=\frac{1}{2}\left[c_{1}^{(2,0)}+3\left(c_{1}^{(3,0)}+c_{2}^{(3,0)}\right)H+3\left(3c_{1}^{(4,0)}+2c_{2}^{(4,0)}+3c_{3}^{(4,0)}\right)H^{2}\right], \label{calG0_xpl}
	\end{equation}
	\begin{equation}
	\mathcal{G}_{1}\left(t\right)=\frac{1}{2}\left[c_{1}^{(2,1)}-\left(c_{1}^{(3,1)}-2c_{2}^{(3,1)}-3c_{3}^{(3,1)}\right)H\right],
	\end{equation}
	\begin{equation}
	\mathcal{G}_{2}\left(t\right)=\frac{1}{2}c_{1}^{(2,2)},
	\end{equation}
	\begin{eqnarray}
	\mathcal{W}_{0}\left(t\right) & = & \frac{1}{4}\Big[2c_{1}^{(0,2)}+\partial_{t}c_{3}^{(1,2)}\nonumber \\
	&  & \quad+\left(3c_{3}^{(1,2)}+6c_{4}^{(1,2)}+2\partial_{t}c_{6}^{(2,2)}+3\partial_{t}c_{8}^{(2,2)}\right)H\nonumber \\
	&  & \quad+\left(4c_{6}^{(2,2)}+6c_{7}^{(2,2)}+9c_{8}^{(2,2)}+18c_{9}^{(2,2)}\right)H^{2}\nonumber \\
	&  & \quad+\left(2c_{6}^{(2,2)}+3c_{8}^{(2,2)}\right)\dot{H}\Big],
	\end{eqnarray}
	\begin{equation}
	\mathcal{W}_{1}\left(t\right)=\frac{1}{4}\partial_{t}\left(c_{1}^{(1,3)}+c_{2}^{(1,3)}\right),
	\end{equation}
	\begin{equation}
	\mathcal{W}_{2}\left(t\right)=-\frac{1}{2}c_{3}^{(0,4)}.  \label{calW2_xpl}
	\end{equation}
GR only contains terms proportional to $\dot{\gamma}_{ij}\dot{\gamma}^{ij}$ and $\gamma_{ij}\Delta\gamma^{ij}$. Other terms in Eq. (\ref{S2T_model}) arise due to the modification of gravity. The term proportional to $\dot{\gamma}_{ij}\Delta\dot{\gamma}^{ij}$
was considered in Refs. \cite{Deruelle:2012xv,Yajima:2015xva}. The two parity-violating terms in (\ref{S2T_model}), i.e., $\epsilon^{ijk}\dot{\gamma}_{li}\partial_{j}\dot{\gamma}_{k}^{l}$ and $\epsilon^{ijk}\gamma_{li}\Delta\partial_{j}\gamma_{k}^{l}$, are considered in Ref. \cite{Nishizawa:2018srh} (see also Ref. \cite{Creminelli:2014wna}).
Note for the term $\epsilon_{ijk}\partial^{i}\dot{\gamma}_{l}^{j}\dot{\gamma}^{kl}$,
Ref. \cite{Nishizawa:2018srh} considered the contribution from $c_{1}^{(2,1)}$ (i.e., $\varepsilon_{ijk}K_{l}^{i}\nabla^{j}K^{kl}$ in the action) only.

From Eq. (\ref{S2T_model}) and the various coefficients $\mathcal{G}_{n}$ and $\mathcal{W}_{n}$ given in Eqs. (\ref{calG0_xpl})-(\ref{calW2_xpl}), there are 21 terms in the original action (\ref{model}) (with 21 free coefficients $c_{1}^{(2,0)}$, $c_{1}^{(3,0)}$ etc.) that contribute to the propagation of linear gravitational waves.
According to Eq. (\ref{WnGn}), in order to make both polarization modes propagate in the speed of light, i.e., $c_{\mathrm{T}}^{(+2)}=c_{\mathrm{T}}^{(-2)}=1$, we must require that $\mathcal{G}_{0}= \mathcal{W}_{0}$, $\mathcal{G}_{1}= \mathcal{W}_{1}$ and $\mathcal{G}_{2}= \mathcal{W}_{2}$.
Moreover, these should be satisfied with any value of $H(t)$, or in other words, they should be stable against the variation of $H$.
With these requirements, we get 7 constraints for the 21 coefficients:
	\begin{equation}
	c_{1}^{(2,0)}-c_{1}^{(0,2)}-\frac{1}{2}\partial_{t}c_{3}^{(1,2)}=0, \label{cond_1}
	\end{equation}
	\begin{equation}
	6c_{1}^{(3,0)}+6c_{2}^{(3,0)}-3c_{3}^{(1,2)}-6c_{4}^{(1,2)}-2\partial_{t}c_{6}^{(2,2)}-3\partial_{t}c_{8}^{(2,2)}=0,
	\end{equation}
	\begin{equation}
	18c_{1}^{(4,0)}+12c_{2}^{(4,0)}+18c_{3}^{(4,0)}-4c_{6}^{(2,2)}-6c_{7}^{(2,2)}-9c_{8}^{(2,2)}-18c_{9}^{(2,2)}=0,
	\end{equation}
	\begin{equation}
	2c_{6}^{(2,2)}+3c_{8}^{(2,2)}=0,
	\end{equation}
	\begin{equation}
	c_{1}^{(2,1)}-\frac{1}{2}\partial_{t}\left(c_{1}^{(1,3)}+c_{2}^{(1,3)}\right)=0,
	\end{equation}
	\begin{equation}
	c_{1}^{(3,1)}-2c_{2}^{(3,1)}-3c_{3}^{(3,1)}=0,
	\end{equation}
	\begin{equation}
	c_{1}^{(2,2)}+c_{3}^{(0,4)}=0. \label{cond_7}
	\end{equation}
From Eqs. (\ref{cond_1})-(\ref{cond_7}), we may solve 7 coefficients to be:
	\begin{eqnarray}
	c_{1}^{(0,2)} & = & c_{1}^{(2,0)}-\frac{1}{2}\partial_{t}c_{3}^{(1,2)}, \label{sol_c021}\\
	c_{1}^{(2,1)} & = & \frac{1}{2}\partial_{t}\left(c_{1}^{(1,3)}+c_{2}^{(1,3)}\right),\\
	c_{4}^{(1,2)} & = & c_{1}^{(3,0)}+c_{2}^{(3,0)}-\frac{1}{2}c_{3}^{(1,2)},\\
	c_{3}^{(3,1)} & = & \frac{1}{3}\left(c_{1}^{(3,1)}-2c_{2}^{(3,1)}\right),\\
	c_{8}^{(2,2)} & = & -\frac{2}{3}c_{6}^{(2,2)},\\
	c_{9}^{(2,2)} & = & \frac{1}{9}\left(9c_{1}^{(4,0)}+6c_{2}^{(4,0)}+9c_{3}^{(4,0)}+c_{6}^{(2,2)}-3c_{7}^{(2,2)}\right),\\
	c_{3}^{(0,4)} & = & -c_{1}^{(2,2)}. \label{sol_c043}
	\end{eqnarray}
The other 14 coefficients are left undetermined.

After plugging Eqs. (\ref{sol_c021})-(\ref{sol_c043}) into Eq. (\ref{model}), and rearranging terms according to the independent coefficients, the action that satisfies $c_{\mathrm{T}}=1$ is given by
	\begin{equation}
	S_{c_{\mathrm{T} = 1}}=\int\!\mathrm{d}t\mathrm{d}^{3}x\,N\sqrt{h}\left(L^{(0)}+L^{(1)}+L^{(2)}+\tilde{L}^{(3)}+\tilde{L}^{(4)}\right), \label{S_cTeq1}
	\end{equation}
where $L^{(0)}$ and $L^{(1)}$ are the same in (\ref{L0})-(\ref{L1}), which do not contribute to the gravitational waves,
	\begin{equation}
	\tilde{L}^{(2)}=c_{1}^{(2,0)}\left(K_{ij}K^{ij}+R\right)+c_{2}^{(2,0)}K^{2},
	\end{equation}
and
	\begin{eqnarray}
	\tilde{L}^{(3)} & = & c_{1}^{(3,0)}\left(K_{ij}K^{jk}K_{k}^{i}+RK\right)+c_{2}^{(3,0)}\left(K_{ij}K^{ij}+R\right)K+c_{3}^{(3,0)}K^{3}\nonumber \\
	&  & +c_{1}^{(1,2)}\nabla^{i}\nabla^{j}K_{ij}+c_{2}^{(1,2)}\nabla^{2}K+c_{3}^{(1,2)}G^{ij}K_{ij}-\frac{1}{2N}\partial_{t}c_{3}^{(1,2)}R,
	\end{eqnarray}
and
	\begin{eqnarray}
	\tilde{L}^{(4)} & = & c_{1}^{(4,0)}\left(K_{ij}K^{jk}K_{k}^{i}+RK\right)K+c_{2}^{(4,0)}\left(\left(K_{ij}K^{ij}\right)^{2}+\frac{2}{3}RK^{2}\right)+c_{3}^{(4,0)}\left(K_{ij}K^{ij}+R\right)K^{2}+c_{4}^{(4,0)}K^{4}\nonumber \\
	&  & +c_{1}^{(3,1)}\varepsilon_{ijk}\left(\nabla_{m}K_{n}^{i}K^{jm}K^{kn}+\frac{1}{3}\nabla^{i}K_{l}^{j}K^{kl}K\right)+c_{2}^{(3,1)}\varepsilon_{ijk}\left(\nabla^{i}K_{m}^{j}K_{n}^{k}K^{mn}-\frac{2}{3}\nabla^{i}K_{l}^{j}K^{kl}K\right)\nonumber \\
	&  & +c_{1}^{(2,2)}\left(\nabla_{k}K_{ij}\nabla^{k}K^{ij}-R_{ij}R^{ij}\right)+c_{2}^{(2,2)}\nabla_{i}K_{jk}\nabla^{k}K^{ij}+c_{3}^{(2,2)}\nabla_{i}K^{ij}\nabla_{k}K_{j}^{k}+c_{4}^{(2,2)}\nabla_{i}K^{ij}\nabla_{j}K\nonumber \\
	&  & +c_{5}^{(2,2)}\nabla_{i}K\nabla^{i}K+c_{6}^{(2,2)}R_{ij}\left(K_{k}^{i}K^{jk}-\frac{2}{3}K^{ij}K+\frac{1}{9}h^{ij}K^{2}\right)+c_{7}^{(2,2)}R\left(K_{ij}K^{ij}-\frac{1}{3}K^{2}\right)\nonumber \\
	&  & +c_{1}^{(1,3)}\varepsilon_{ijk}R^{il}\nabla^{j}K_{l}^{k}+c_{2}^{(1,3)}\varepsilon_{ijk}\nabla^{i}R_{l}^{j}K^{kl}+\frac{1}{2N}\partial_{t}\left(c_{1}^{(1,3)}+c_{2}^{(1,3)}\right)\varepsilon_{ijk}K_{l}^{i}\nabla^{j}K^{kl}\nonumber \\
	&  & +c_{1}^{(0,4)}\nabla^{i}\nabla^{j}R_{ij}+c_{2}^{(0,4)}\nabla^{2}R+c_{4}^{(0,4)}R^{2}. \label{Ltld_4}
	\end{eqnarray}
We conclude that the action (\ref{S_cTeq1}) represents a large class of gravity theories respecting the spatial symmetry, in which both polarization modes of the gravitational waves propagate in the speed of light in the cosmological background.
Please note that spatial derivatives of $N$ can be added into Eq.(\ref{S_cTeq1}), which do not affect the linear GWs.

\subsubsection{On Horndeski theory with $c_{\mathrm{T}} = 1$}

As a simple application of our result, let us consider the Horndeski theory, of which the Lagrangian in the unitary gauge is given in (\ref{LH_ug}). The conditions (\ref{cond_1})-(\ref{cond_7}) simply reduce to
	\begin{equation}
	b-d_{1}+\frac{1}{N}\partial_{t}a_{1}=0,\label{Hdsk_c1}
	\end{equation}
and
	\begin{equation}
	c=0.\label{Hdsk_c2}
	\end{equation}
(\ref{Hdsk_c1}) implies
	\begin{equation}
	b-d_{1}+\frac{1}{N}\partial_{t}a_{1} =  N\frac{\partial G_{4}}{\partial N}+\frac{1}{N^{2}}\frac{\partial G_{5}}{\partial t}=0. \label{Hdsk_c1_xp}
	\end{equation}
On the other hand, from (\ref{Hdsk_c}), (\ref{Hdsk_c2}) implies $G_{5} = G_{5}(t)$. 
Using (\ref{Hdsk_c1_xp}), we may rewrite $b$ to be
	\begin{equation}
	b=G_{4}-\frac{1}{2N^{2}}\frac{\partial G_{5}}{\partial t},
	\end{equation}
and thus
	\begin{equation}
	\frac{\partial b}{\partial N}=\frac{\partial G_{4}}{\partial N}+\frac{1}{N^{3}}\frac{\partial G_{5}}{\partial t}\equiv0,
	\end{equation}
which implies $b=b(t)$. Finally, after some manipulations, we arrive at the conclusion that under the requirement $c_{\mathrm{T}}=1$, the Horndeski action in the unitary gauge reduces to be
	\begin{equation}
	S_{\mathrm{H},c_\mathrm{T}=1}^{\mathrm{(u.g.)}}=\int\mathrm{d}t\mathrm{d}^{3}x\,N\sqrt{h}\left[b\left(t\right)\left(K_{ij}K^{ij}-K^{2}+R\right)+a_{0}\,K+d_{0}\right],
	\end{equation}
where $a_0$ and $d_0$ can be general functions of $t$ and $N$.
The above analysis is consistent with the previous results (e.g.) \cite{Creminelli:2017sry}.

\subsubsection{On parity-violating gravity}

In (\ref{Ltld_4}), we recognize four parity-violating terms that preserve $c_{\mathrm{T}} = 1$ for both polarization modes:
	\begin{eqnarray}
	\mathcal{O}_{1} & = & c_{1}\,\varepsilon_{ijk}\left(\nabla_{m}K_{n}^{i}K^{jm}K^{kn}+\frac{1}{3}\nabla^{i}K_{l}^{j}K^{kl}K\right), \label{pv_O1}\\
	\mathcal{O}_{2} & = & c_{2}\,\varepsilon_{ijk}\left(\nabla^{i}K_{m}^{j}K_{n}^{k}K^{mn}-\frac{2}{3}\nabla^{i}K_{l}^{j}K^{kl}K\right),\\
	\mathcal{O}_{3} & = & \varepsilon_{ijk}\left(c_{3}R^{il}\nabla^{j}K_{l}^{k}+\frac{1}{2N}\partial_{t}c_{3}K_{l}^{i}\nabla^{j}K^{kl}\right),\\
	\mathcal{O}_{4} & = & \varepsilon_{ijk}\left(c_{4}\nabla^{i}R_{l}^{j}K^{kl}+\frac{1}{2N}\partial_{t}c_{4}K_{l}^{i}\nabla^{j}K^{kl}\right), \label{pv_O4}
	\end{eqnarray}
where $c_{1},\cdots,c_{4}$ are general functions of $t$ and $N$, although generally they may also depend on spatial derivatives of lapse function $N$.
At this point, it is interesting to note that the CS gravity (\ref{CS_fP_ug}) corresponds to the special choice of parameters with
	\begin{eqnarray}
	c_{1} & = & -8f,\\
	c_{2} & = & +8f,\\
	c_{3} & = & -16f,\\
	c_{4} & = & 0.
	\end{eqnarray}
This explains the fact that the speed of gravitational waves in CS gravity is not modified.
One finding in this work is that there exist more general parity-violating terms that have this property.

\subsection{Effective mass in the UV}

Due to the presence of spatial derivatives of $K_{ij}$ in the theory, the appearance of $k$-dependence in $\mathcal{G}^{(s)}$ introduces interesting features for the propagation of the gravitational waves.
Here we briefly mention one interesting feature by revealing that the gravitational waves may become effectively massive in the short-wavelength limit (or ultra-violate limit).

First let us consider a simpler case where both $\mathcal{G}^{(s)}$ and $\mathcal{W}^{(s)}$ are up to $k^{2}$, which imply that $\mathcal{G}_{2}\neq 0$ and $\mathcal{W}_{2}\neq 0$. In this case Eq. (\ref{cT2_k4}) becomes
	\begin{equation}
	\left(c_{\mathrm{T}}^{(s)}\right)^{2}=\frac{\mathcal{W}_{0}\left(t\right)+\mathcal{W}_{1}\left(t\right)\frac{s}{2}\frac{k}{a}+\mathcal{W}_{2}\left(t\right)\frac{k^{2}}{a^{2}}}{\mathcal{G}_{0}\left(t\right)+\mathcal{G}_{1}\left(t\right)\frac{s}{2}\frac{k}{a}+\mathcal{G}_{2}\left(t\right)\frac{k^{2}}{a^{2}}}.
	\end{equation}
In the short-wavelength limit (i.e., $k/a\rightarrow\infty$) we have
	\begin{eqnarray}
	\left(c_{\mathrm{T}}^{(s)}\right)^{2} & \rightarrow & \frac{\mathcal{W}_{2}}{\mathcal{G}_{2}}+\frac{\mathcal{G}_{2}\mathcal{W}_{1}-\mathcal{G}_{1}\mathcal{W}_{2}}{\mathcal{G}_{2}^{2}}\frac{s}{2}\frac{a}{k}\nonumber \\
	&  & +\frac{\mathcal{G}_{1}^{2}\mathcal{W}_{2}-\mathcal{G}_{2}\left(\mathcal{G}_{1}\mathcal{W}_{1}+\mathcal{G}_{0}\mathcal{W}_{2}\right)+\mathcal{G}_{2}^{2}\mathcal{W}_{0}}{\mathcal{G}_{2}^{3}}\frac{a^{2}}{k^{2}}+\mathcal{O}\left(k^{-3}\right).
	\end{eqnarray}
If we further require
	\begin{equation}
	\mathcal{G}_{2}\mathcal{W}_{1}-\mathcal{G}_{1}\mathcal{W}_{2}=0,
	\end{equation}
we get
	\begin{equation}
	k^{2}\left(c_{\mathrm{T}}^{(s)}\right)^{2}\rightarrow c_{\mathrm{UV}}^{2}k^{2}+m_{\mathrm{UV}}^{2}+\mathcal{O}\left(k^{-1}\right).
	\end{equation}
with
	\begin{eqnarray}
	c_{\mathrm{UV}}^{2} & = & \frac{\mathcal{W}_{2}}{\mathcal{G}_{2}},\\
	m_{\mathrm{UV}}^{2} & = & a^{2}\frac{\mathcal{G}_{2}\mathcal{W}_{0}-\mathcal{G}_{0}\mathcal{W}_{2}}{\mathcal{G}_{2}^{2}}.
	\end{eqnarray}
$m_{\mathrm{UV}}$ is thus the effective mass for the gravitational waves in the UV limit.

For the full-version of Eq. (\ref{cT2_k4}), in the limit of $k/a\rightarrow\infty$ we have 
	\begin{eqnarray}
	\left(c_{\mathrm{T}}^{(s)}\right)^{2} & \rightarrow & \frac{\mathcal{W}_{4}}{\mathcal{G}_{4}}+\frac{\left(\mathcal{G}_{4}\mathcal{W}_{3}-\mathcal{G}_{3}\mathcal{W}_{4}\right)}{\mathcal{G}_{4}^{2}}\frac{s}{2}\frac{a}{k}\\
	&  & +\frac{\mathcal{G}_{3}^{2}\mathcal{W}_{4}-\mathcal{G}_{4}\left(\mathcal{G}_{3}\mathcal{W}_{3}+\mathcal{G}_{2}\mathcal{W}_{4}\right)+\mathcal{G}_{4}^{2}\mathcal{W}_{2}}{\mathcal{G}_{4}^{3}}\frac{a^{2}}{k^{2}}+\mathcal{O}\left(k^{-3}\right).
	\end{eqnarray}
If we further require
	\begin{equation}
	\mathcal{G}_{4}\mathcal{W}_{3}-\mathcal{G}_{3}\mathcal{W}_{4}=0,
	\end{equation}
we get
	\begin{equation}
	k^{2}\left(c_{\mathrm{T}}^{(s)}\right)^{2}\rightarrow c_{\mathrm{UV}}^{2}k^{2}+m_{\mathrm{UV}}^{2}+\mathcal{O}\left(k^{-1}\right),
	\end{equation}
with
	\begin{eqnarray}
	c_{\mathrm{UV}}^{2} & = & \frac{\mathcal{W}_{4}}{\mathcal{G}_{4}},\\
	m_{\mathrm{UV}}^{2} & = & a^{2}\frac{\mathcal{G}_{4}\mathcal{W}_{2}-\mathcal{G}_{2}\mathcal{W}_{4}}{\mathcal{G}_{4}^{2}}.
	\end{eqnarray}
Again, in the UV limit the gravitational waves acquire an effective mass term $m_{\mathrm{UV}}$.
Of course, the above results are derived in the UV limit, which implies that the effective mass must be of the sub-leading order. Nevertheless, it is interesting to explore this effect, which is a distinctive feature of the spatial derivatives of the extrinsic curvature.

%%%%%%%%%%
\section{Conclusion} \label{sec:con}

In this work we investigated the propagation of the gravitational waves in a cosmological background.
Based on the framework of spatially covariant gravity proposed in Refs. \cite{Gao:2014soa,Gao:2014fra}, we derived the general quadratic action for the gravitational waves (\ref{S2T_gen}).
Two types of terms are systematically introduced: the spatial derivatives of the extrinsic curvature and the parity-violating terms.
From Eq. (\ref{S2T_gen}) and the resulting equation of motion Eq. (\ref{eom_gammas}), spatial derivatives of the extrinsic curvature will modify the dispersion relation in a different manner and make the Planck mass running rate $k$-dependent, which thus modify the amplitude of the gravitational waves in a $k$-dependent manner.
This, on the other hand, raises the possibility of keeping the propagation speed $c_{\mathrm{T}}$ unchanged while modifying the Planck mass running rate only.

We paid special attention to the propagation speed of the gravitational waves. The detection of GW170817 and its electromagnetic counterpart \cite{TheLIGOScientific:2017qsa,Monitor:2017mdv} implies the coincidence of the propagation speeds of the GWs and of the light. 
In this work, we tried to explore the possibility of having $c_{\mathrm{T}} = 1$ within the framework of spatially covariant gravity.
We found that it is possible to make the two circular polarization modes propagate in the same speed, even in the presence of parity-violating terms. In particular, we found a large class of spatially covariant gravity theories (\ref{S_cTeq1}) with parity-violation (\ref{pv_O1})-(\ref{pv_O4}), in which both polarization modes of the GWs propagate in the speed of light.
Previously, this property was known for the CS gravity. In this work we shown that there are more general parity-violating gravity theories having this property, or in other words, surviving under the restriction $c_{\mathrm{T}}=1$.
Our results imply that, although the parameter space of the covariant scalar-tensor theories is heavily restricted, the spatially covariant gravity may provide us more possibilities in light of the propagation of the gravitational waves.

% % % % % % % % % % % % % %

\acknowledgments

We would like to thank Yi-Fu Cai, Hai-Shan Liu, Yu-Xiao Liu, Hong L\"{u}, Anzhong Wang, Yi Wang, Wen Zhao, Tao Zhu for helpful discussions.
We are grateful to Wen Zhao for valuable comments on a preliminary draft of this work.
X.G. would like to thank the Institute of Theoretical Physics of the Lanzhou University, the Institute for Advanced Physics and Mathematics of the Zhejiang University of Technology, the Center for Gravitation and Cosmology of the Yangzhou University, the Department of Astronomy of the University of Science and Technology of China, The Jockey Club Institute for Advanced Study of the Hong Kong University of Science and Technology for hospitality, where the work was in the final stage during his stay.
X.G. was supported by the Chinese National Youth Thousand Talents Program (No. 71000-41180003) and by the SYSU start-up funding.

% % % % % % % % % %
\appendix

\section{Parity-violating theories} \label{app:pv} 

Some interesting parity-violating gravity theories that are healthy in the unitary gauge were found in Ref. \cite{Crisostomi:2017ugk}. Here we reformulate their results.

One class of terms considered in Ref. \cite{Crisostomi:2017ugk} are quadratic in the Riemann tensor and involve only the first derivative of the scalar field.  
There are 4 independent terms:
	\begin{eqnarray}
	\mathcal{O}_{1} & = & \varepsilon^{\mu\nu\rho\sigma}\,{}^{4}\! R_{\rho\sigma\alpha\beta}\,{}^{4}\! R_{\mu\nu\phantom{\alpha}\lambda}^{\phantom{\mu\nu}\alpha}\nabla^{\beta}\phi\nabla^{\lambda}\phi,\\
	\mathcal{O}_{2} & = & \varepsilon^{\mu\nu\rho\sigma}\,{}^{4}\! R_{\rho\sigma\alpha\beta}\,{}^{4}\! R_{\phantom{\beta}\nu}^{\beta}\nabla^{\alpha}\phi\nabla_{\mu}\phi,\\
	\mathcal{O}_{3} & = & \varepsilon^{\mu\nu\rho\sigma}\,{}^{4}\! R_{\rho\sigma\alpha\beta}\,{}^{4}\! R_{\mu\lambda}^{\phantom{\mu g}\alpha\beta}\nabla_{\nu}\phi \nabla^{\lambda}\phi,\\
	\mathcal{O}_{4} & = & \varepsilon^{\mu\nu\rho\sigma}\,{}^{4}\! R_{\rho\sigma\alpha\beta}\,{}^{4}\! R_{\phantom{\alpha\beta}\mu\nu}^{\alpha\beta}\nabla_{\lambda}\phi \nabla^{\lambda}\phi.
	\end{eqnarray}
If we consider the linear combination
	\begin{equation}
	S=\int\mathrm{d}^{4}x\sqrt{-g}\left(\sum_{n=1}^{4}A_{n}\mathcal{O}_{n}\right),
	\end{equation}
where $A_{1},\cdots,A_{4}$ are general functions of $\phi$ and $X\equiv -\frac{1}{2}(\partial\phi)^2$, the coefficients must satisfy
	\begin{equation}
	4A_{1}+A_{2}+2A_{3}+8A_{4}=0,
	\end{equation}
in order to make the theory to be healthy in the unitary gauge \cite{Crisostomi:2017ugk}.
This can be understood more transparently by observing that in the unitary gauge, there are 3 independent combinations of $\mathcal{O}_{1},\cdots,\mathcal{O}_{4}$ in which the ``dangerous'' term $\pounds_{\bm{n}}K_{ij}$ exactly drops out:
\begin{equation}
\mathcal{O}_{1}-\frac{1}{2}\mathcal{O}_{4},\qquad\mathcal{O}_{2}-\frac{1}{8}\mathcal{O}_{4},\qquad\mathcal{O}_{3}-\frac{1}{4}\mathcal{O}_{4},
\end{equation}
which are nothing but $\mathcal{L}_{\mathrm{A},1},\mathcal{L}_{\mathrm{A},2},\mathcal{L}_{\mathrm{A},3}$ in Eqs. (\ref{pv_LA1}), (\ref{pv_LA2}) and (\ref{pv_LA3}), respectively.

Another class of terms considered in Ref. \cite{Crisostomi:2017ugk} are linear in the Riemann tensor and quadratic in the second derivative of the scalar field. There are 5 independent terms
	\begin{eqnarray}
	\mathcal{O}_{1} & = & \varepsilon^{\mu\nu\rho\sigma}\,{}^{4}\! R_{\rho\sigma\alpha\beta}\nabla^{\alpha}\nabla_{\mu}\phi\nabla^{\beta}\nabla_{\nu}\phi\nabla_{\lambda}\phi\nabla^{\lambda}\phi,\\
	\mathcal{O}_{2} & = & \varepsilon^{\mu\nu\rho\sigma}\,{}^{4}\! R_{\rho\sigma\alpha\beta}\nabla^{\alpha}\nabla_{\mu}\phi\nabla^{\lambda}\nabla_{\nu}\phi\nabla^{\beta}\phi\nabla_{\lambda}\phi,\\
	\mathcal{O}_{3} & = & \varepsilon^{\mu\nu\rho\sigma}\,{}^{4}\! R_{\rho\alpha\beta\lambda}\nabla^{\beta}\nabla_{\mu}\phi\nabla^{\lambda}\nabla_{\nu}\phi\nabla^{\alpha}\phi\nabla_{\sigma}\phi,\\
	\mathcal{O}_{4} & = & \varepsilon^{\mu\nu\rho\sigma}\,{}^{4}\! R_{\rho\sigma\alpha\beta}\nabla^{\alpha}\nabla_{\mu}\phi\nabla^{\beta}\nabla_{\lambda}\phi\nabla_{\nu}\phi\nabla^{\lambda}\phi,\\
	\mathcal{O}_{5} & = & \varepsilon^{\mu\nu\rho\sigma}\,{}^{4}\! R_{\sigma\alpha}\nabla^{\alpha}\nabla_{\mu}\phi\nabla^{\beta}\nabla_{\nu}\phi\nabla_{\rho}\phi\nabla_{\beta}\phi,
	\end{eqnarray}
up to the quadratic order in the first derivative $\nabla_{\mu}\phi$.
If we consider the combination
	\begin{equation}
	S=\int\mathrm{d}^{4}x\sqrt{-g}\left(\sum_{n=1}^{5}A_{n}\mathcal{O}_{n}\right),
	\end{equation}
where $A_{1},\cdots,A_{5}$ are general functions of $\phi$ and $X\equiv -\frac{1}{2}(\partial\phi)^2$, the coefficients must satisfy
	\begin{eqnarray}
	4A_{1}+2A_{2}+2A_{3}-A_{5} & = & 0,\\
	2A_{1}+A_{2}+A_{4} & = & 0,
	\end{eqnarray}
in order to make the theory to be healthy in the unitary gauge, where the ``dangerous'' terms $\pounds_{\bm{n}}K_{ij}$ and $\pounds_{\bm{n}}N$ exactly get cancelled.
There are thus 3 combinations
	\begin{equation}
	\mathcal{O}_{1}-2\mathcal{O}_{4}+4\mathcal{O}_{5},\qquad\mathcal{O}_{2}-\mathcal{O}_{4}+2\mathcal{O}_{5},\qquad\mathcal{O}_{3}+2\mathcal{O}_{5},
	\end{equation}
which are exactly $\mathcal{L}_{\mathrm{C},1},\mathcal{L}_{\mathrm{C},2},\mathcal{L}_{\mathrm{C},3}$ in Refs. (\ref{pv_LC1}), (\ref{pv_LC2}) and (\ref{pv_LC3}), respectively.

% % % % % % % % % % % % % %


\end{document}